\documentclass[twocolumn,letterpaper,aps,prc,superscriptaddress,showpacs,nofootinbib,floatfix]{revtex4}

\usepackage{graphicx}	

\newcommand{\pt}{{p_T}}
\newcommand{\jpsi}{{\rm{J}/\psi}}
\newcommand{\chic}{{\chi_c}}
\newcommand{\psiprime}{{\psi'}}
\newcommand{\raa}{{R_{\rm AA}}}
\newcommand{\avrg}[1] {\langle #1\rangle}

\newcommand{\snn}{{\sqrt{s_{\rm NN}}}}

\newcommand{\pp}{{p+p}}
\newcommand{\cucu}{{{\rm Cu}+{\rm Cu}}}

\renewcommand{\aa}{{{\rm A}+{\rm A}}}

\newcommand{\auau}{{\rm Au}+{\rm Au}}
\newcommand{\ncol}{{N_{\rm coll}}}
\newcommand{\npart}{{N_{\rm part}}}

\newcommand{\refrad}{{r_{\rm ref}}}
\newcommand{\bfrefrad}{{\bf r_{\rm \bf ref}}}
\newcommand{\zbbc}{{z_{\rm BBC}}}

\newcommand{\zfit}{{z_{\rm fit}}}
\newcommand{\bfzfit}{{\bf z_{\rm \bf fit}}}
\newcommand{\pdt}{{p\delta\theta}}
\newcommand{\bfpdt}{{\mathbf p\delta\theta}}
\newcommand{\dz}{{\delta z}}
\newcommand{\bfdz}{{\mathbf\delta z}}

\newcommand{\bfnhitmutr}{{N_{\rm \bf hits}^{\rm \bf MuTr}}}

\newcommand{\bfnhitmuid}{{N_{\rm \bf hits}^{\rm \bf MuID}}}

\newcommand{\nf}{{N_F}}
\newcommand{\nc}{{N_C}}
\renewcommand{\ni}{{N_I}}

\newcommand{\ccbar}{{c\overline{c}}}

\begin{document}


\title{Nuclear-Modification Factor for Open-Heavy-Flavor Production 
at Forward Rapidity in Cu$+$Cu Collisions at $\sqrt{s_{\rm NN}}=200$~GeV}
   
\newcommand{\abilene}{Abilene Christian University, Abilene, Texas 79699, USA}
\newcommand{\banaras}{Department of Physics, Banaras Hindu University, Varanasi 221005, India}
\newcommand{\bnlcoll}{Collider-Accelerator Department, Brookhaven National Laboratory, Upton, New York 11973-5000, USA}
\newcommand{\bnlphys}{Brookhaven National Laboratory, Upton, New York 11973-5000, USA}
\newcommand{\caucr}{University of California - Riverside, Riverside, California 92521, USA}
\newcommand{\charlesczech}{Charles University, Ovocn\'{y} trh 5, Praha 1, 116 36, Prague, Czech Republic}
\newcommand{\ciae}{Science and Technology on Nuclear Data Laboratory, China Institute of Atomic Energy, Beijing 102413, P.~R.~China}
\newcommand{\cns}{Center for Nuclear Study, Graduate School of Science, University of Tokyo, 7-3-1 Hongo, Bunkyo, Tokyo 113-0033, Japan}
\newcommand{\colorado}{University of Colorado, Boulder, Colorado 80309, USA}
\newcommand{\columbia}{Columbia University, New York, New York 10027 and Nevis Laboratories, Irvington, New York 10533, USA}
\newcommand{\czechtech}{Czech Technical University, Zikova 4, 166 36 Prague 6, Czech Republic}
\newcommand{\dapnia}{Dapnia, CEA Saclay, F-91191, Gif-sur-Yvette, France}
\newcommand{\debrecen}{Debrecen University, H-4010 Debrecen, Egyetem t{\'e}r 1, Hungary}
\newcommand{\elte}{ELTE, E{\"o}tv{\"o}s Lor{\'a}nd University, H - 1117 Budapest, P{\'a}zm{\'a}ny P. s. 1/A, Hungary}
\newcommand{\fit}{Florida Institute of Technology, Melbourne, Florida 32901, USA}
\newcommand{\fsu}{Florida State University, Tallahassee, Florida 32306, USA}
\newcommand{\gsu}{Georgia State University, Atlanta, Georgia 30303, USA}
\newcommand{\hiroshima}{Hiroshima University, Kagamiyama, Higashi-Hiroshima 739-8526, Japan}
\newcommand{\ihepprot}{IHEP Protvino, State Research Center of Russian Federation, Institute for High Energy Physics, Protvino, 142281, Russia}
\newcommand{\illuiuc}{University of Illinois at Urbana-Champaign, Urbana, Illinois 61801, USA}
\newcommand{\inrras}{Institute for Nuclear Research of the Russian Academy of Sciences, prospekt 60-letiya Oktyabrya 7a, Moscow 117312, Russia}
\newcommand{\instpasczech}{Institute of Physics, Academy of Sciences of the Czech Republic, Na Slovance 2, 182 21 Prague 8, Czech Republic}
\newcommand{\isu}{Iowa State University, Ames, Iowa 50011, USA}
\newcommand{\jinrdubna}{Joint Institute for Nuclear Research, 141980 Dubna, Moscow Region, Russia}
\newcommand{\kek}{KEK, High Energy Accelerator Research Organization, Tsukuba, Ibaraki 305-0801, Japan}
\newcommand{\korea}{Korea University, Seoul, 136-701, Korea}
\newcommand{\kurchatov}{Russian Research Center ``Kurchatov Institute", Moscow, 123098 Russia}
\newcommand{\kyoto}{Kyoto University, Kyoto 606-8502, Japan}
\newcommand{\labllr}{Laboratoire Leprince-Ringuet, Ecole Polytechnique, CNRS-IN2P3, Route de Saclay, F-91128, Palaiseau, France}
\newcommand{\lawllnl}{Lawrence Livermore National Laboratory, Livermore, California 94550, USA}
\newcommand{\losalamos}{Los Alamos National Laboratory, Los Alamos, New Mexico 87545, USA}
\newcommand{\lpc}{LPC, Universit{\'e} Blaise Pascal, CNRS-IN2P3, Clermont-Fd, 63177 Aubiere Cedex, France}
\newcommand{\lund}{Department of Physics, Lund University, Box 118, SE-221 00 Lund, Sweden}
\newcommand{\muenster}{Institut f\"ur Kernphysik, University of Muenster, D-48149 Muenster, Germany}
\newcommand{\myongji}{Myongji University, Yongin, Kyonggido 449-728, Korea}
\newcommand{\nagasaki}{Nagasaki Institute of Applied Science, Nagasaki-shi, Nagasaki 851-0193, Japan}
\newcommand{\newmex}{University of New Mexico, Albuquerque, New Mexico 87131, USA }
\newcommand{\nmsu}{New Mexico State University, Las Cruces, New Mexico 88003, USA}
\newcommand{\ornl}{Oak Ridge National Laboratory, Oak Ridge, Tennessee 37831, USA}
\newcommand{\orsay}{IPN-Orsay, Universite Paris Sud, CNRS-IN2P3, BP1, F-91406, Orsay, France}
\newcommand{\peking}{Peking University, Beijing 100871, P.~R.~China}
\newcommand{\pnpi}{PNPI, Petersburg Nuclear Physics Institute, Gatchina, Leningrad region, 188300, Russia}
\newcommand{\riken}{RIKEN Nishina Center for Accelerator-Based Science, Wako, Saitama 351-0198, Japan}
\newcommand{\rikjrbrc}{RIKEN BNL Research Center, Brookhaven National Laboratory, Upton, New York 11973-5000, USA}
\newcommand{\rikkyo}{Physics Department, Rikkyo University, 3-34-1 Nishi-Ikebukuro, Toshima, Tokyo 171-8501, Japan}
\newcommand{\saispbstu}{Saint Petersburg State Polytechnic University, St. Petersburg, 195251 Russia}
\newcommand{\saopaulo}{Universidade de S{\~a}o Paulo, Instituto de F\'{\i}sica, Caixa Postal 66318, S{\~a}o Paulo CEP05315-970, Brazil}
\newcommand{\seoulnat}{Seoul National University, Seoul, Korea}
\newcommand{\stonybrkc}{Chemistry Department, Stony Brook University, SUNY, Stony Brook, New York 11794-3400, USA}
\newcommand{\stonycrkp}{Department of Physics and Astronomy, Stony Brook University, SUNY, Stony Brook, New York 11794-3400, USA}
\newcommand{\subatech}{SUBATECH (Ecole des Mines de Nantes, CNRS-IN2P3, Universit{\'e} de Nantes) BP 20722 - 44307, Nantes, France}
\newcommand{\tenn}{University of Tennessee, Knoxville, Tennessee 37996, USA}
\newcommand{\titech}{Department of Physics, Tokyo Institute of Technology, Oh-okayama, Meguro, Tokyo 152-8551, Japan}
\newcommand{\tsukuba}{Institute of Physics, University of Tsukuba, Tsukuba, Ibaraki 305, Japan}
\newcommand{\vandy}{Vanderbilt University, Nashville, Tennessee 37235, USA}
\newcommand{\waseda}{Waseda University, Advanced Research Institute for Science and Engineering, 17 Kikui-cho, Shinjuku-ku, Tokyo 162-0044, Japan}
\newcommand{\weizmann}{Weizmann Institute, Rehovot 76100, Israel}
\newcommand{\wigner}{Institute for Particle and Nuclear Physics, Wigner Research Centre for Physics, Hungarian Academy of Sciences (Wigner RCP, RMKI) H-1525 Budapest 114, POBox 49, Budapest, Hungary}
\newcommand{\yonsei}{Yonsei University, IPAP, Seoul 120-749, Korea}
\affiliation{\abilene}
\affiliation{\banaras}
\affiliation{\bnlcoll}
\affiliation{\bnlphys}
\affiliation{\caucr}
\affiliation{\charlesczech}
\affiliation{\ciae}
\affiliation{\cns}
\affiliation{\colorado}
\affiliation{\columbia}
\affiliation{\czechtech}
\affiliation{\dapnia}
\affiliation{\debrecen}
\affiliation{\elte}
\affiliation{\fit}
\affiliation{\fsu}
\affiliation{\gsu}
\affiliation{\hiroshima}
\affiliation{\ihepprot}
\affiliation{\illuiuc}
\affiliation{\inrras}
\affiliation{\instpasczech}
\affiliation{\isu}
\affiliation{\jinrdubna}
\affiliation{\kek}
\affiliation{\korea}
\affiliation{\kurchatov}
\affiliation{\kyoto}
\affiliation{\labllr}
\affiliation{\lawllnl}
\affiliation{\losalamos}
\affiliation{\lpc}
\affiliation{\lund}
\affiliation{\muenster}
\affiliation{\myongji}
\affiliation{\nagasaki}
\affiliation{\newmex}
\affiliation{\nmsu}
\affiliation{\ornl}
\affiliation{\orsay}
\affiliation{\peking}
\affiliation{\pnpi}
\affiliation{\riken}
\affiliation{\rikjrbrc}
\affiliation{\rikkyo}
\affiliation{\saispbstu}
\affiliation{\saopaulo}
\affiliation{\seoulnat}
\affiliation{\stonybrkc}
\affiliation{\stonycrkp}
\affiliation{\subatech}
\affiliation{\tenn}
\affiliation{\titech}
\affiliation{\tsukuba}
\affiliation{\vandy}
\affiliation{\waseda}
\affiliation{\weizmann}
\affiliation{\wigner}
\affiliation{\yonsei}
\author{A.~Adare} \affiliation{\colorado}
\author{S.~Afanasiev} \affiliation{\jinrdubna}
\author{C.~Aidala} \affiliation{\columbia}
\author{N.N.~Ajitanand} \affiliation{\stonybrkc}
\author{Y.~Akiba} \affiliation{\riken} \affiliation{\rikjrbrc}
\author{H.~Al-Bataineh} \affiliation{\nmsu}
\author{J.~Alexander} \affiliation{\stonybrkc}
\author{K.~Aoki} \affiliation{\kyoto} \affiliation{\riken}
\author{L.~Aphecetche} \affiliation{\subatech}
\author{R.~Armendariz} \affiliation{\nmsu}
% ...
\collaboration{PHENIX Collaboration} \noaffiliation
   
\date{\today}


\begin{abstract}

{\bf Background:} Heavy-flavor production in $p$+$p$~collisions is a 
good test of perturbative-quantum-chromodynamics (pQCD) 
calculations.  Modification of heavy-flavor production in heavy-ion 
collisions relative to binary-collision scaling from $p$+$p$ results, 
quantified with the nuclear-modification factor ($R_{\rm AA}$), 
provides information on both cold- and hot-nuclear-matter effects. 
Midrapidity heavy-flavor $R_{\rm AA}$ measurements at RHIC have 
challenged parton-energy-loss models and resulted in upper limits 
on the viscosity/entropy ratio that are near the quantum lower 
bound. Such measurements have not been made in the forward-rapidity 
region.

{\bf Purpose:} Determine transverse-momentum, $\pt$ spectra and the 
corresponding $R_{\rm AA}$ for muons from 
heavy-flavor mesons decay in $p$+$p$ and Cu$+$Cu collisions at 
$\sqrt{s_{NN}}=200$\,GeV and $y = 1.65$.

{\bf Method:} Results are obtained using the semi-leptonic decay of 
heavy-flavor mesons into negative muons. The PHENIX muon-arm 
spectrometers measure the $p_T$ spectra of inclusive muon 
candidates.  Backgrounds, primarily due to light hadrons, are 
determined with a Monte-Carlo calculation using a set of input 
hadron distributions tuned to match measured-hadron distributions 
in the same detector and statistically subtracted.

{\bf Results:} The charm-production cross section in 
$p$+$p$ collisions at $\sqrt{s}=200$\,GeV, integrated over $\pt$
and in the rapidity range $1.4<y<1.9$ is found to be 
$d\sigma_{c\bar{c}}/dy = 0.139\pm 0.029\ 
{\rm (stat)\,}^{+0.051}_{-0.058}\ {\rm (syst)}$~mb. This result is 
consistent with a perturbative fixed-order-plus-next-to-leading-log 
(FONLL) calculation within scale uncertainties and is also 
consistent with expectations based on the corresponding 
midrapidity charm-production cross section measured by PHENIX. The 
$R_{\rm AA}$ for heavy-flavor muons in Cu$+$Cu  
collisions is measured in three centrality intervals for 
$1<\pt<4$~GeV/$c$. Suppression relative to binary-collision scaling 
($R_{\rm AA} < 1$) increases with centrality.

{\bf Conclusions:} Within experimental and theoretical 
uncertainties, the measured heavy-flavor yield in $p$+$p$ 
collisions is consistent with state-of-the-art pQCD calculations.  
Suppression in central Cu$+$Cu collisions suggests the presence 
of significant cold-nuclear-matter effects and final-state 
energy loss.

\end{abstract}



	
\maketitle

\section{Introduction}\label{sec:introduction}

Understanding the energy loss mechanism for partons moving through 
the hot dense partonic matter produced in heavy-ion collisions at 
the Relativistic Heavy Ion Collider (RHIC) and the Large Hadron 
Collider (LHC) is a key priority in the field of heavy-ion 
collision physics~\cite{tannenbaum:2006,d'enterria:2010}. 
Production of heavy quarks in heavy-ion collisions can serve as an 
important tool for better understanding properties of the dense 
matter created in such collisions. In particular, because of their 
large mass, heavy quarks are almost exclusively produced in the 
early stages of heavy-ion collisions and can therefore serve as a 
probe of the subsequently created medium. The large mass scale 
associated with the production of heavy quarks also allows one to 
test perturbative Quantum Chromodynamics (pQCD) based theoretical 
models describing high energy collisions.

Recent measurements of heavy-quark production in heavy-ion 
collisions~\cite{Adler:2005xv,Adare:2007hf,Adare:2011hf} exhibit a 
suppression, which is larger than expected and not easily reconciled 
with early theoretical 
predictions~\cite{Djordjevic:2004nq,Armesto:2005zy}. In these 
calculations the dominant energy loss mechanism for fast partons is 
gluon bremsstrahlung~\cite{Gyulassy:1993hr,Baier:1994bd}. In this 
context, it was predicted that heavy quarks would lose less energy 
than light quarks due to the so-called {\em dead-cone 
effect}~\cite{Dokshitzer:2001zm}. The disagreement between this 
prediction and experimental results led to a consideration of 
alternative in-medium parton energy loss mechanisms, assumed 
earlier to have a small effect on heavy quarks compared to 
radiative energy loss. In particular, it was suggested that heavy 
quarks can lose a significant amount of their energy through 
elastic collisions with in-medium partons (collisional energy loss 
mechanism)~\cite{Mustafa:2005,Moore_Teaney:2005,VanHees:2005}, 
especially in the intermediate transverse momentum range 
($\pt\approx 3--8$\,GeV/$c$) in which most of the RHIC open heavy 
flavor measurements are performed. Additional mechanisms for 
in-medium energy loss for heavy quarks have also been 
suggested~\cite{Adil:2006ra,Rapp:2009my}.  Despite recent progress, 
still needed is a universal theoretical framework describing precisely 
the production of heavy quarks and their subsequent interactions 
with the partonic medium created in heavy-ion collisions.  Also needed 
are accurate measurements of heavy-quark 
production in heavy-ion collisions, which are critical to test and 
constrain the theoretical predictions.

Hidden-heavy-flavor ($\jpsi$) production has also been extensively 
measured in heavy-ion collisions~\cite{Adare:2006ns,Adare:2008sh}. 
The production of $\jpsi$ mesons is expected to be affected by the 
formation of a quark-gluon plasma due to the interplay of several 
competing mechanisms, including suppression due to a color 
screening mechanism similar to the Debye screening in 
QED~\cite{Satz:1986} and enhancement due to the coalescence of 
uncorrelated $\ccbar$ pairs from the hot 
medium~\cite{Andronic:2003,Svetitsky:1988,Thews:2005}. The 
magnitude of such an enhancement depends strongly on the production 
cross section of open-heavy flavor in heavy-ion collisions, 
measurements of which are therefore essential to the interpretation 
of heavy quarkonia results.

A well-established observable for quantifying medium effects in 
heavy-ion collisions is the nuclear-modification factor, 
$\raa$:

\begin{equation}\label{eq:RAA}
\raa=\frac{1}{\ncol}\frac{\sigma_{\rm AA}}{\sigma_{pp}},
\end{equation}
where $\sigma_{\rm AA}$ and $\sigma_{pp}$ are the invariant 
cross sections for a given process in $\aa$ collisions and 
$\pp$ collisions, respectively, and $\ncol$ is the average number 
of nucleon-nucleon collisions in the $\aa$ collision, evaluated using 
a simple geometrical description of the $\rm A$ 
nucleus~\cite{Miller:2007ri}.

For processes that are sufficiently hard (characterized by large 
energy transfer), $\raa$ is expected to be 
equal to unity in the absence of nuclear effects. A value smaller 
(larger) than unity indicates suppression (enhancement) of the 
observed yield in $\aa$ collisions relative to expectations based 
on $\pp$ collision results and binary-collision scaling.

Open-heavy-flavor production has been measured by the PHENIX 
experiment at midrapidity ($|\eta|<0.35$)~\cite{Adler:2005xv}. 
This paper presents the measurement of open-heavy-flavor production 
at forward rapidity ($1.4<|\eta|<1.9$) in $\cucu$ and $\pp$ 
collisions, and the resulting $\raa$, using 
negatively-charged muons from the semi-leptonic decay of 
open-heavy-flavor mesons.

The paper is organized as follows: Section~\ref{sec:detector} 
presents a short overview of the PHENIX detector subsystems 
relevant to these measurements followed by a description of the 
data sets and track selection criteria. Section~\ref{sec:analysis} 
presents a detailed description of the methodology for measuring 
the invariant cross section in $\pp$ collisions and $\raa$
in $\cucu$ collisions for muons from heavy-flavor-meson decays. 
Results are presented in 
Section~\ref{sec:results} and compared to existing measurements as 
well as theoretical predictions in Section~\ref{sec:discussion}.

\section{Experimental setup and data sets}\label{sec:detector}

\subsection{The PHENIX Experiment}

\begin{figure}
\includegraphics[width=1.0\linewidth]{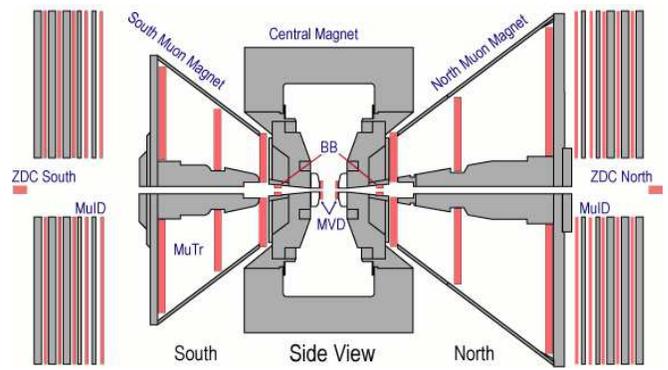}
\caption{(color online) 
Side view of the PHENIX muon detectors (2005).}
\label{fig:phenix}
\end{figure}

The PHENIX experiment is equipped with two muon 
spectrometers~\cite{Adcox:2003zm}, shown in 
Fig.~\ref{fig:phenix}, that provide pion rejection at the level 
of $2.5\times10^{-4}$ in the pseudorapidity range $-1.2<\eta<-2.2$ 
(south muon arm) and $1.2<\eta<2.4$ (north muon arm) over the full 
azimuth. Each muon arm is located behind a thick copper and iron 
absorber and comprises three stations of cathode strip chambers 
(the Muon Tracker, or MuTr), surrounded by a radial magnetic field, 
and five "gaps" (numbered 0--4) consisting of a plane of steel 
absorber and a plane of Iarrocci tubes (the Muon Identifier, or 
MuID). The MuTr measures the momentum of charged particles by 
tracking their motion in the surrounding magnetic field. Matching 
the momentum of the particles reconstructed in the MuTr to the 
penetration depth of the particle in the MuID (that is, the last 
MuID gap a given particle reaches) is the primary tool used to 
identify muons with respect to the residual hadronic background. 
Measured muons must penetrate 8 to 11 interaction lengths in total 
to reach the last gap of the MuID. This corresponds to a reduction 
of the muon longitudinal momentum (along the beam axis) of 
$\delta p_z=2.3 (2.45)$\,GeV/$c$ in the south (north) muon arm. The 
MuID is also used in the online data acquisition to trigger on 
collisions that contain one or more muon candidates.

Also used in this analysis are the Beam-Beam Counters 
(BBC)~\cite{Allen:2003a}, which comprise two arrays of 64 quartz 
\v{C}erenkov detectors that surround the beam, one on each side of the 
interaction point. The BBCs measure charged particles produced 
during the collision in the pseudorapidity range $3<|\eta|<3.9$ 
and determine the collision's start-time, vertex longitudinal 
position, and centrality (in $\cucu$ collisions). The BBCs also 
provide the minimum bias trigger.

\subsection{Data Sets}

Two data sets, recorded in 2005, are used in this analysis: $\pp$ 
collisions and $\cucu$ collisions at a center of mass energy per 
nucleon-nucleon collision of $\sqrt{s_{\rm NN}}=200$\,GeV.

The $\pp$ data used for this analysis have been recorded using two 
{\em muon enriched} triggers, in coincidence with the Minimum Bias 
(MB) trigger, which requires at least one hit in each of the BBCs 
and covers approximately 55\% of the total $\pp$ inelastic cross 
section. These two muon triggers rely on the information recorded 
in the MuID. The first ({\em Deep}) trigger requires one or more 
muon candidates to reach the last plane of the MuID (Gap4), whereas 
the second, less strict, ({\em Shallow}) trigger requires one or 
more muon candidates to reach at least the third MuID gap (Gap2). 
The integrated luminosity sampled with these triggers and used for 
this analysis is $44.3$~nb$^{-1}$ (48.7~nb$^{-1}$) for the south 
(north) muon arm.

All $\cucu$ data used for this analysis have been recorded using 
the Minimum Bias trigger described above. For $\cucu$ collisions, 
this trigger covers approximately 94\,\% of the total inelastic 
cross section. The integrated luminosity sampled with this trigger 
and used for this analysis is 0.13~nb$^{-1}$, using a total $\cucu$ 
inelastic cross section seen by the minimum bias trigger 
$\sigma_{\cucu}^{\rm inel}=2.91$~b.


\subsection{Centrality Determination}

The centrality of each $\cucu$ collision is determined by the 
number of hits in the BBCs.  Three centrality bins are used for 
this analysis: $0--20$\%, $20--40$\% and $40--94$\%, where 
$0--20$\% 
represents the most central 20\% of the collisions. For a given 
centrality, the average number of nucleon-nucleon collisions 
($\ncol$) and the average number of participating nucleons 
($\npart$) are estimated using a Glauber 
calculation~\cite{Miller:2007ri} coupled to a model of the BBC 
response. Values of $\ncol$ and $\npart$ for the three centrality 
bins defined above are listed in Table~\ref{table_centrality}.

To ensure that the centrality categories are well defined, 
collisions are required to be within $\pm 30$\,cm of the center of 
the PHENIX detector along the beam axis.

\begin{table}
\caption{\label{table_centrality}Centrality characterization 
variables for $\cucu$ collisions.}
\begin{ruledtabular}
\begin{tabular}{cccc}
centrality&0--20\%&20--40\%&40--94\%\\
\hline
$\ncol$& 151.8 $\pm$ 17.1 & 61.6 $\pm$ 6.6 & 11.23 $\pm$ 1.3\\
$\npart$& 85.9 $\pm$ 2.3 & 45.2 $\pm$ 1.7 & 11.7 $\pm$ 0.6\\
\end{tabular}
\end{ruledtabular}
\end{table}


\subsection{Track Selection}\label{ssec_track_selection}

This section outlines the track-based selection variables. 

\begin{description}

\item [{$\bf \zbbc$}] The event vertex longitudinal position is 
measured by the BBC detector. For low-momentum tracks ($p_T < 
2$\,GeV/$c$) reconstructed in north (south) muon arm we demand $\zbbc 
> 0$ ($\zbbc < 0$). This arm-dependent cut improves the signal to 
background ratio because light hadrons produced during the 
collision have a probability to decay into a muon that increases 
with their distance from the front muon arm absorber, whereas muons 
from short-lived heavy-flavor hadrons have a yield that is 
independent of $\zbbc$ (see also Section~\ref{sssec:tuning}).

\item [{$\bfzfit$}] The vertex longitudinal position of a track 
evaluated using a fit of the track position and momentum measured 
in the MuTr and extrapolated backward through the front absorber 
towards the interaction point, together with the BBC vertex 
measurement.

\item [${\bfnhitmutr}$] The total number of track hits in the 
three MuTr stations. A given track can have up to 16 MuTr hits.

\item [${\bfnhitmuid}$] The total number of track hits in the 
five MuID gaps. A given track can have up to 2 hits in each gap (10 
in total).

\item [${\bfrefrad}$] The distance to the beam axis of the 
track, as reconstructed in the MuID only, when extrapolated 
(backward) to $z=0$ (illustrated in Fig.~\ref{fig:refrad}).

\begin{figure}
\includegraphics[width=1.0\linewidth]{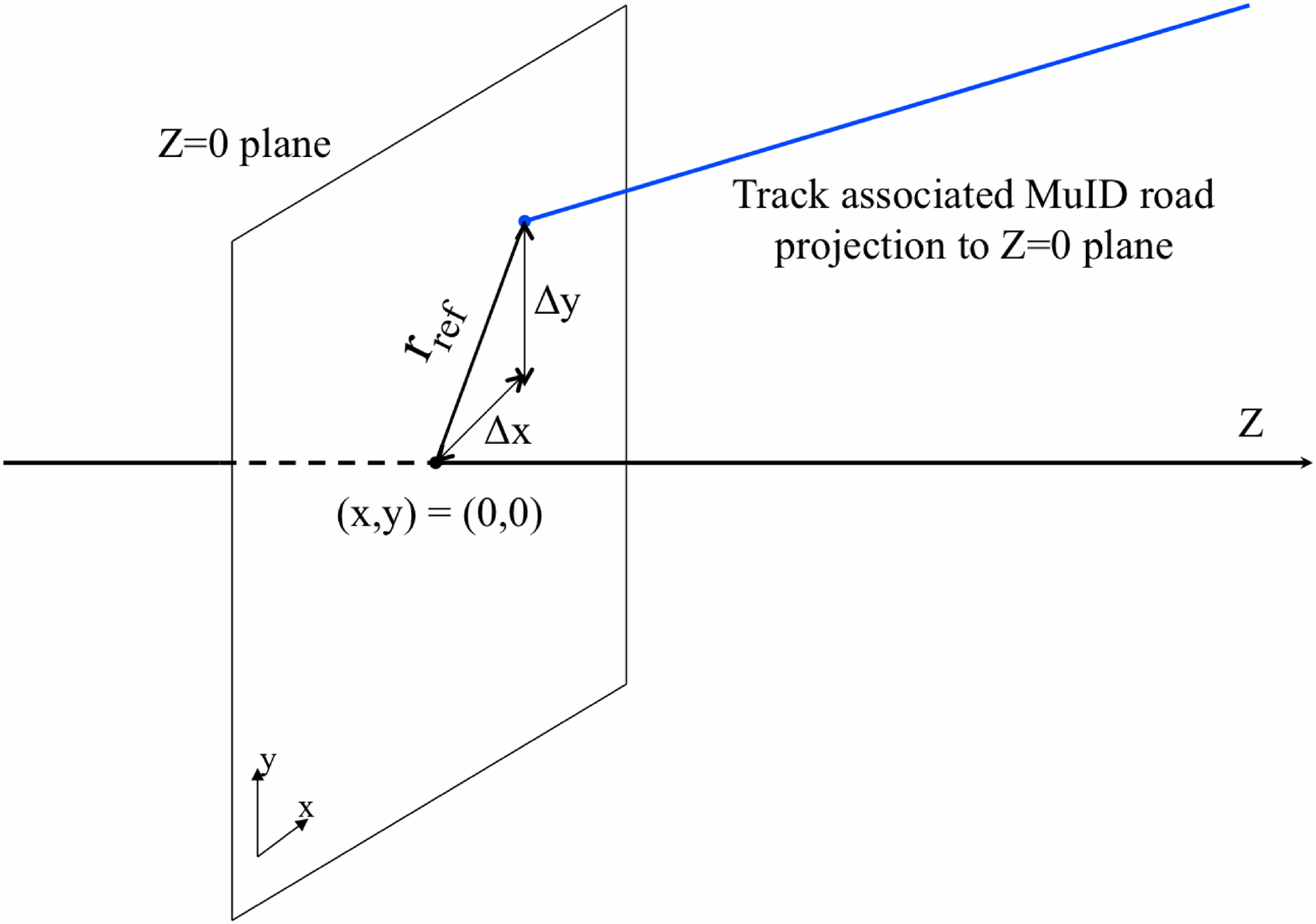}
\caption{(color online)
Schematic representation of $\refrad$ variable.}
\label{fig:refrad}

\includegraphics[width=1.0\linewidth]{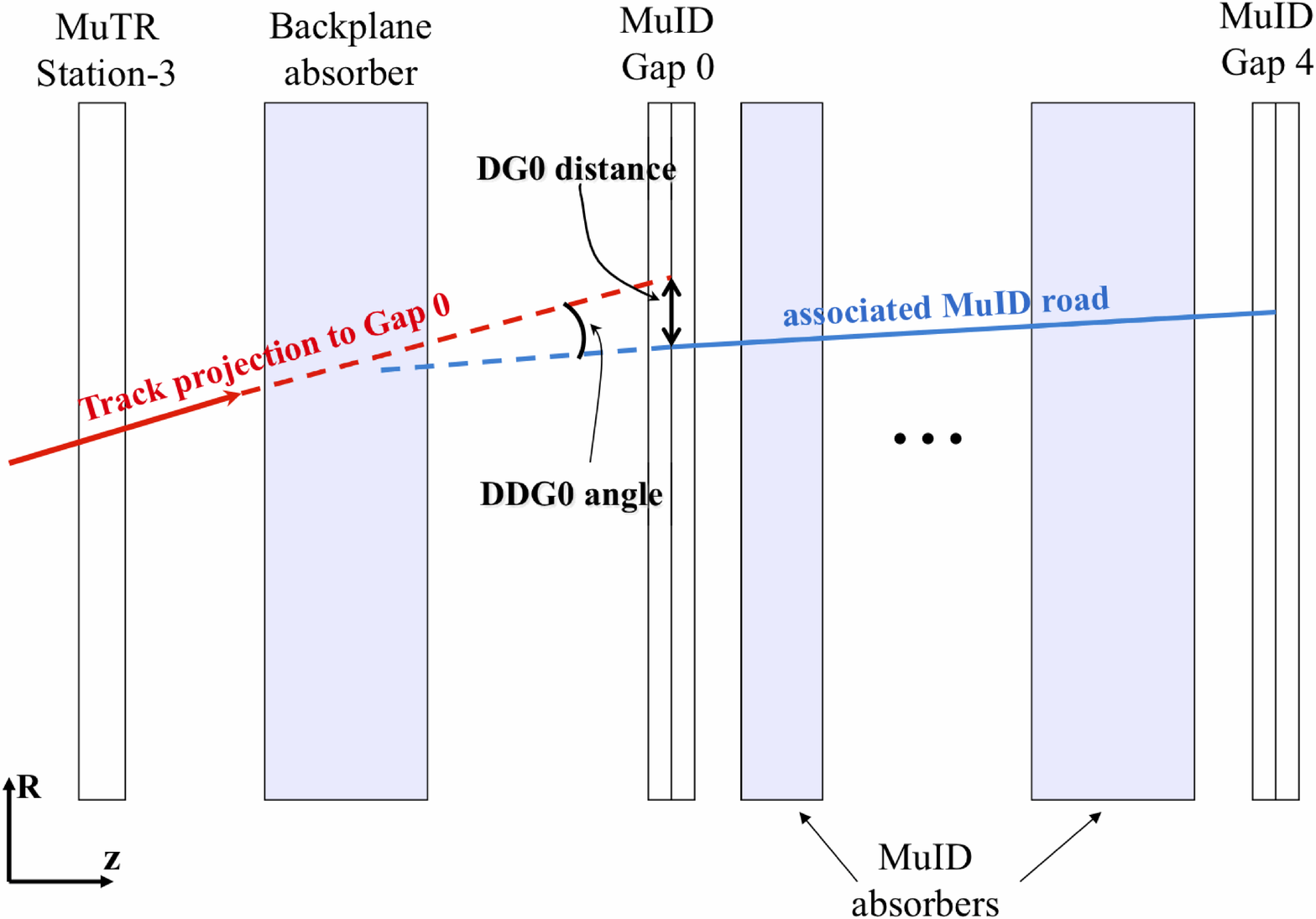}
\caption{(color online)
Schematic representation of track selection variables DG0 
and DDG0.}
\label{fig:dg0}
\end{figure}

\item [Road Slope] The slope of the track, as 
reconstructed in the MuID only, measured at MuID Gap0: 
$\sqrt{(dx/dz)^{2} + (dy/dz)^{2} }$. A cut applied to this variable 
eliminates combinatorial background generated in the high 
hit-density region closest to the beam pipe.

\item [DG0] The distance between the track positions 
calculated in the MuTr and in the MuID, evaluated at the MuID Gap0 
$z$ position (see Fig.~\ref{fig:dg0}).

\item [DDG0] The difference between the track angles 
calculated in the MuTr and in the MuID, evaluated at the MuID Gap0 
$z$ position (see Fig.~\ref{fig:dg0}).

\item [${\bfdz}$] The difference between the event vertex 
longitudinal position reconstructed by the BBC ($\zbbc$) and the 
track longitudinal position provided by the track reconstruction 
algorithm: $\delta z=|\zbbc - \zfit|$.

\item [${\bfpdt}$] the {\em effective} scattering angle of the 
track in the front absorber, $\delta\theta$, scaled by the average 
of the momentum measured at the vertex and at MuTr Station 1: $p$ = 
$(p_{\rm vtx} + p_{\rm st1})/2$, where $\delta\theta$ is given by:   
\begin{equation}
\delta \theta = \cos^{-1}\left(\frac{\overrightarrow{p}_{\rm vtx} \cdot \overrightarrow{p}_{\rm st1} } {{p}_{\rm vtx}.{p}_{\rm st1} } \right). 
\end{equation}
where $\overrightarrow{p}_{\rm st1}$ is the momentum vector 
measured at Station 1 and $\overrightarrow{p}_{\rm vtx}$ is the 
momentum vector at the vertex. For a given track, $\delta\theta$ 
essentially measures the track deflection in the front absorber due 
mostly to multiple scattering and radiative energy loss, but also 
to the magnetic field upstream of station 1. This deflection is 
expected to be inversely proportional to the track total momentum. 
Scaling the scattering angle $\delta\theta$ by the track momentum 
therefore ensures that the $\pdt$ distribution is approximately 
Gaussian with a constant width for all $\pt$ bins.

\end{description}

Cut values applied to these variables are, in some cases, $\pt$-, 
species- and/or centrality-dependent. Within a given $\pt$, species 
and centrality bin, the same cut values are applied to both Monte 
Carlo simulations and real data.

Even after all cuts are applied to select {\em good quality} muon 
candidates, there remains a small contamination of 
misreconstructed tracks caused by:

\begin{itemize}

\item Accidental combinations of hits in the muon tracker that do 
not correspond to a real particle.


\item Tracks arising from interactions between the beam and 
residual gas in the beam pipe or between the beam and beamline 
components.

\end{itemize}

These misreconstructed tracks, later denoted $\nf$, are not 
completely reproduced by experimental simulations and must be 
estimated and properly subtracted from the inclusive muon sample to 
evaluate the amount of muons from heavy-flavor decay. The method by 
which $\nf$ is estimated is based on the distributions of the 
$\pdt$ and $\delta z$ variables and is described in more detail in 
Section~\ref{ssec_fake_tracks}.

Note:  positive muons are not used in this analysis due to a poorer 
signal/background ratio resulting from the fact that both 
anti-protons and negative kaons are more strongly suppressed by the 
MuTr front absorbers than their positive counterparts. The rapidity 
interval used for this measurement is smaller than the rapidity 
coverage of the PHENIX muon spectrometers ($1.2<|\eta|<2.2$) to 
reduce uncertainties in the acceptance calculation.

\section{Method for the Measurement of Heavy-Flavor Muons}
\label{sec:analysis}

\subsection{\label{ssec_analysis_common}Overview}

The methodology used to measure heavy-flavor muon (i.e., muons from 
heavy-flavor meson decay) production in $\pp$ and $\cucu$ 
collisions is described in this section. This analysis is a 
refinement of techniques originally developed in 
\cite{Adler:2006yu,Iraklis_thesis,Donnys_thesis}.

For both $\pp$ and $\cucu$ collisions the double differential heavy 
flavor muon invariant yield is defined by:

\begin{equation}\label{eq:smyield}
\frac{d^{2}N^{\mu}}{2 \pi \pt d\pt d\eta} 
= \frac{1}{2\pi\pt\Delta\pt \Delta\eta}\frac{\ni-\nc-\nf}
{N_{\rm evt}\epsilon_{\rm BBC}^{c\overline{c}\rightarrow \mu}A\epsilon}
\end{equation}
where $\ni$ is the total number of muon candidates in the bin, 
consisting of the tracks that reach the last gap of the MuID (Gap4) 
and pass all track selection criteria; $\nf$ is the estimated 
number of misreconstructed tracks that pass the track selection 
cuts accidentally (Section~\ref{ssec_fake_tracks}); $\nc$ is the 
number of tracks corresponding to the irreducible hadronic 
background, as determined using a hadron cocktail approach 
(Section~\ref{ssec_hadron_cocktail}); $N_{\rm evt}$ is the number 
of events, $A\epsilon$ is the detector acceptance and efficiency 
correction (Section~\ref{ssec_acc_eff}), and $\epsilon_{\rm 
BBC}^{c\overline{c}\rightarrow \mu}$ is the BBC trigger efficiency 
for events in which a heavy-flavor muon at forward rapidity is 
present. This efficiency amounts to 79\% (100\%) in $\pp$ ($\cucu$) 
collisions.

The $\pp$ and $\cucu$ invariant yields determined with 
Eq.~\ref{eq:smyield} can be used directly to determine the 
heavy-flavor muon $\raa$ 
(Eq.~\ref{eq:RAA}). However, in order to minimize the 
systematic uncertainty associated with the estimate of the hadronic 
background by canceling the part of this uncertainty that is 
correlated between the $\pp$ and the $\cucu$ analyses, $\raa$ is 
calculated separately for a given $i^{th}$ version of the 
Monte-Carlo simulation of hadron cocktail used in the estimate of $\nc$:
\begin{equation}\label{eq:modRAA}
R^{i}_{\rm AA} = \frac{1}{\ncol}\left(\frac{d^2N_{\cucu}/d\pt d\eta}
{d^2N_{\pp}/d\pt d\eta}\right)^{i}
\end{equation}
The final value for $\raa$ is then determined by taking the mean of 
the values obtained for the different cocktails, each weighted by 
its ability to reproduce measured data, as discussed in 
Section~\ref{ssec_central_point_determination}.

\subsection{Contamination from Misreconstructed Tracks}\label{ssec_fake_tracks}

\begin{figure}
\includegraphics[width=1.0\linewidth]{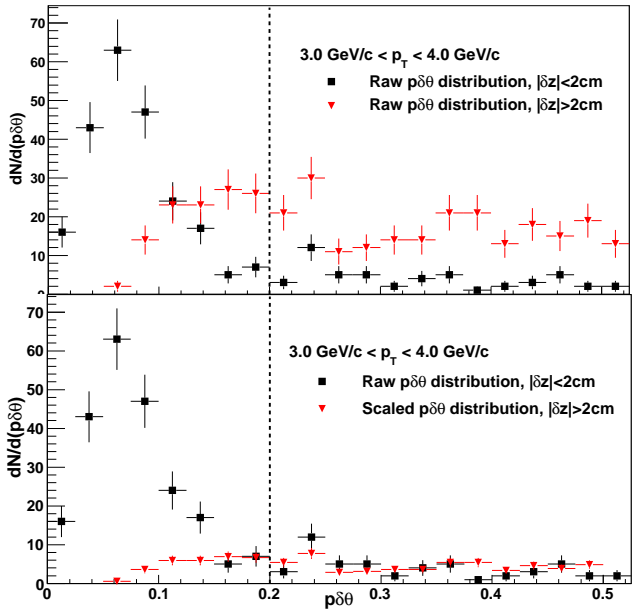}
\caption{(color online)
$\pdt$ distributions for north arm inclusive muon 
candidates, $3 < \pt < 4$\,GeV/$c$. The top panel compares the 
distribution inside (black squares) and outside (red triangles) the 
$\dz$ cut. The bottom panel compares the same distributions, but 
the distribution outside the $\dz$ cut (red triangles) is 
normalized to the distribution inside the $\dz$ cut (black squares) 
in the region $\pdt > \pdt_{\rm max}$. In both panels, the vertical 
dashed line corresponds to $\pdt_{\rm max}$.}
\label{fig:pdtheta_scaled}
\end{figure}

$\nf$, the number of misreconstructed tracks that accidentally 
pass all track quality cuts, is estimated using the $\pdt$ 
distribution inside and outside of the $\dz$ cut defined in 
Section~\ref{ssec_track_selection}. These two distributions are 
shown in the top panel of Fig.~\ref{fig:pdtheta_scaled}. The 
distribution inside the $\dz$ cut (black squares) shows two 
contributions: a peak at $\pdt=0.05$\,rad$\cdot$GeV/$c$, 
corresponding to the expected multiple scattering of muons in the 
front absorber, and a tail out to large values of $\pdt$. In the 
distribution outside the $\dz$ cut (red triangles), the signal peak 
has disappeared, and only the tail remains. Note that the tail 
extends below the $\pdt$ cut; this is the $\nf$ contribution. Using 
the fact that the shape of this tail appears to be the same on both 
sides of the $\dz$ cut, one can estimate $\nf$ using:

\begin{equation}
\nf = \alpha \nf'
\label{eq_nf}
\end{equation}
where $\nf'$ is the number of tracks with $\pdt < \pdt_{\rm max}$ but 
$\dz > \dz_{\rm max}$, and $\alpha$ normalizes the tails of the two 
distributions above the $\pdt$ cut:
\begin{equation}
\alpha=\frac{N(\pdt>\pdt_{\rm max},\dz<\dz_{\rm max})}{N(\pdt>\pdt_{\rm max},\dz>\dz_{\rm max})}
\label{eq_nf_scale}
\end{equation}

The bottom panel of Fig.~\ref{fig:pdtheta_scaled} shows the 
$\pdt$ distribution inside the $\dz$ cut (black squares, identical 
to the corresponding distribution in the top panel) and the 
distribution outside the $\dz$ cut (red triangles from the top 
panel) after scaling by $\alpha$ (Eq.~\ref{eq_nf_scale}).

Using Equations~\ref{eq_nf} and \ref{eq_nf_scale}, it is found that 
$\nf$ amounts to less than $1\%$ of the inclusive muon sample in 
the lowest $\pt$ bin ($1<\pt<1.5$~GeV/$c$) and increases with $\pt$ 
up to about $5\%$ for the highest $\pt$ bins. Uncertainties on 
these estimations are negligible in the final results.

\subsection{Hadron Cocktail}\label{ssec_hadron_cocktail}

Charged pions and kaons are the largest source of particles in the 
PHENIX muon arms. Other species ($p$, $\bar{p}$, $K^{0}_{s}$, 
$K^{0}_{L}$) have small but nonzero contributions. Altogether, 
these light hadrons constitute the main background source for the 
measurement of muons from heavy-flavor meson decay.

One can define three contributions to this background, depending on 
how the particles enter the muon spectrometer:

\begin{description}

\item [Decay muons] - light hadrons that decay into muons 
before reaching the first absorber material. Since these particles 
enter the spectrometer as muons, a fraction of them also penetrate 
all the absorber layers of the MuID and enter the pool of inclusive 
muon candidates.

\item [Punch-through hadrons] - hadrons produced at the 
collision vertex that do not decay, but penetrate all MuID absorber 
layers, thus also being (incorrectly) identified as muons.

\item [Decay-in-MuTr] - hadrons produced at the collision 
vertex that penetrate the muon arm front absorber and decay into a 
muon inside the MuTr tracking volume, with the decay muon then 
passing through the rest of the MuTr and the MuID. Most such 
particles are simply not reconstructed because of the decay angle 
between the primary hadron and the decay muon. However, some can be 
reconstructed, usually with an incorrect momentum assigned to the 
track. Due to the exponential $p_T$ distribution, even a small 
number of such tracks can form a significant background at high 
$p_T$, but for the $p_T$ range in this analysis this contribution 
is small.

\end{description}

\begin{figure}
 \includegraphics[width=1.0\linewidth]{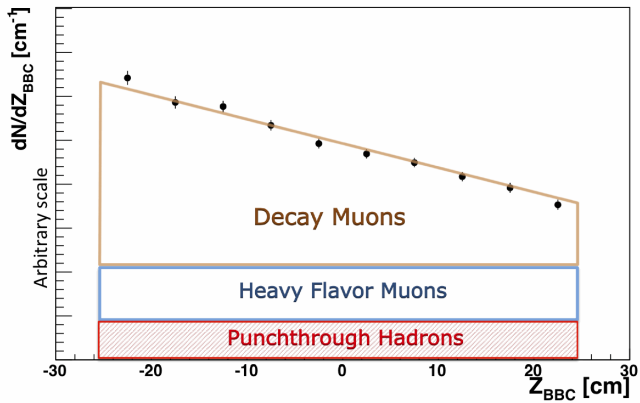}
\caption{(color online)
Vertex $z$ distribution of muon candidates reconstructed in north 
($z>0$) MuID Gap4, relative to the event vertex $z$ distribution 
(black circles). The vertex $z$ dependencies of the various 
contributions to the inclusive muon spectra are represented 
schematically as colored boxes.} 
\label{fig:inclusive_content}
\end{figure}

While decay muons can not be distinguished from punch-through 
hadrons and heavy-flavor muons on an event-by-event basis, their 
production exhibits a strong vertex dependence, as illustrated 
in Fig.~\ref{fig:inclusive_content}. This feature plays a key role 
in constraining heavy-flavor background 
(Section~\ref{sssec:tuning}).

A series of Monte Carlo simulations (``hadron cocktail packages") 
are used to estimate the overall background due to light hadron 
sources. The construction of a given hadronic cocktail package 
involves the following steps:

\begin{enumerate}

\item Generate a primary hadron sample based on parameterized $\pt$ 
and $y$ distributions (Section~\ref{sssec_generation}).

\item Propagate these hadrons through the muon spectrometer using 
the complete {\sc geant}3~\cite{geant3} PHENIX simulation. Each hadron 
cocktail package uses one of the two hadron shower codes provided 
by {\sc geant}3: {\sc g-fluka} or {\sc gheisha} with a scaled value of the 
hadron-Iron interaction cross section 
(Section~\ref{sssec:packages}).

\item For the $\cucu$ analysis the simulated hadrons are then 
embedded in real events in order to account for deterioration of 
the reconstructed track quality due to high hit multiplicity.

\item Reconstruct the resulting particles using the same 
reconstruction code and track quality cuts used in the real data 
analysis. (Section~\ref{ssec_track_selection}).

\item Tune (that is, re-weight) the input $\pt$ distributions (from 
step 1) to match hadron distributions measured in the muon arm 
(Section~\ref{sssec:tuning}).

\end{enumerate}

\subsubsection{Input Particle Distributions}\label{sssec_generation}

Particle distributions required as input to the hadron cocktail 
have not been measured over the required $y$ and $\pt$ range at 
RHIC energies. We therefore use a combination of data from PHENIX, 
BRAHMS and STAR, together with Next-to-Leading Order (NLO) pQCD 
calculations to derive realistic parameterizations of these 
distributions. An exact match to actual distributions is not 
necessary since the input distributions are re-weighted to match 
measured hadron distributions before being used to generate 
estimates of $\nc$ (Section~\ref{sssec:tuning}).

We start with the $\pi^0$ spectrum in $\pp$ collisions at $y=0$ 
measured by PHENIX~\cite{Adler:2003pi}. This is extrapolated to $y 
= 1.65$ in two steps. First, an overall scale factor is obtained 
from a Gaussian parameterization of the charged pion $dN/dy$ 
distribution measured by BRAHMS~\cite{videbaek_proceedings}. Next, 
the $\pt$ shape is softened using a parameterization of the ratio 
of unidentified hadron $\pt$ spectra measured by BRAHMS at $\eta = 
0$ and $\eta = 1.65$~\cite{Samsets_thesis,Arsene:2004jd}:

\begin{widetext}
\begin{equation}
dN/d\pt(\pi^{\pm},y=1.65) =
dN/d\pt(\pi^{0},y=0) \times \exp(-\frac{1}{2}(1.65/2.25)^{2}) \times (1- (0.1\cdot \pt[{\rm GeV}/c] -1) )
\label{eq:chargedpionyields}
\end{equation}
\end{widetext}

\begin{figure}
 \includegraphics[width=1.0\linewidth]{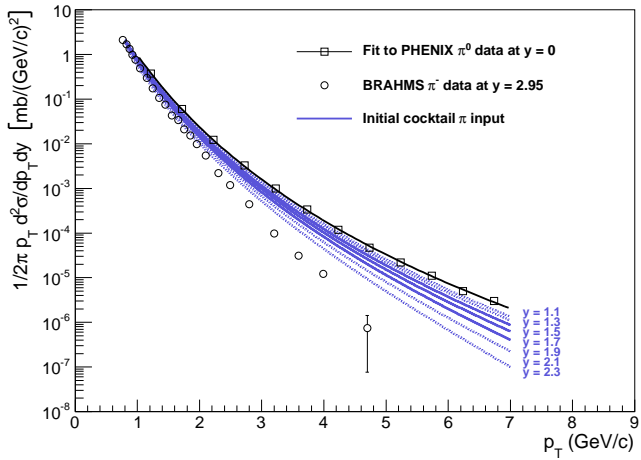}
\caption{(color online)
Pion cross sections as a function of $\pt$ used as initial hadron 
cocktail input, for several rapidity intervals in $[1.0,2.2]$ (blue 
lines) compared to a fit to the PHENIX $\pi^{0}$ data at $y=0$ 
\cite{Adler:2003pi}(black line, open black circles) and BRAHMS 
$\pi^{-}$ data at $y=2.95$ \cite{Arsene:2007jd} (open black 
circles).}
\label{fig:Cinput_pi}
 \end{figure}

Next we extrapolate this spectrum over the range $1.0 \le y \le 
2.4$ using a series of Next-to-Leading Order (NLO) 
calculations~\cite{Vogelsang:private} to obtain the ratio 
$dN/d\pt(\pi^{\pm},y)/dN/d\pt(\pi^{\pm},y=1.65)$. 
Figure~\ref{fig:Cinput_pi} shows a comparison of the hadron 
cocktail input for charged pions compared to 
charged-pion distributions at $y=0$ and $y=2.95$. Spectra for 
other hadron species in the cocktail are obtained by multiplying 
the parameterized pion spectra by parameterizations of measured 
values of hadron-to-pion ratios, as a function of $\pt$.

With 8--11 interaction length of material prior to MuID Gap4, 
approximately 4000 hadrons must be simulated to obtain a single 
hadron reconstructed as a muon. Given this level of rejection, it 
is very CPU intensive to generate a sufficient sample of high $\pt$ 
hadrons using realistic $\pt$ spectra. A standard technique is to 
throw particles with a flat $\pt$ spectrum and then weight them 
with a realistic distribution. However, interactions in the 
absorber in front of the MuTr and decays in the MuTr volume can 
both result in particles being reconstructed with incorrect 
momentum. Due to the steeply falling nature of the $\pt$ spectrum, 
tracks with low momentum and incorrectly reconstructed with a 
higher momentum can have a significant contribution at high $\pt$, 
with respect to properly reconstructed tracks. As a compromise 
designed to ensure statistically robust samples of both tracks with 
{\em initial} high $\pt$ and with {\em misreconstructed} high 
$\pt$, we multiply the realistic $\pt$ distributions by $\pt^2$ to 
form the simulation input $\pt$ distributions, and re-weight the 
output of the simulation by $1/\pt^2$ to recover the initial 
distribution.

The particles in the primary hadron sample used as input to each 
hadron cocktail package are generated as follows:

\begin{itemize}

\item The particle type and rapidity are chosen based on $dN/dy$ 
values obtained by integrating the unweighted $\pt$ distributions 
described above.

\item The particle's transverse momentum is chosen within the range 
$0.8\le\pt\le8$~GeV/$c$ using the $\pt^2$-weighted $\pt$ 
distributions described above.

\item Since the muon spectrometer acceptance shows little 
dependence on the vertex $z$ position, the particle's $z$ origin is 
chosen from a flat distribution over the range $-35\le z\le35$~cm.

\item The particle's azimuthal angle, $\phi$, is chosen from a flat 
distribution over $2\pi$.

\end{itemize}

\subsubsection{Hadron Cocktail Packages}\label{sssec:packages}

Modeling hadron propagation through thick material is known to be 
difficult and neither hadron shower code available in {\sc geant}3 
({\sc g-fluka} and {\sc gheisha}) is able to reproduce measured data in the 
PHENIX muon arms. The approach we have chosen to circumvent this 
issue is to produce a range of background estimates using a set of 
hadron cocktails (referred to as {\em packages}), each of which 
uses one of the {\sc geant} hadron shower codes and a different, 
modified, value of the hadron-Iron interaction cross section. The 
set of background estimates are then combined in a weighted fashion 
to extract central values for production yields, $\raa$, and 
the contribution to the systematic uncertainty on these quantities 
due to the uncertainty in hadron propagation.

Using the default hadron-ion cross section, {\sc fluka} 
simulations 
produce more muon candidates than {\sc gheisha} simulations, therefore 
the {\sc fluka} cross sections are increased relative to the default and 
the {\sc gheisha} cross sections are decreased.  The cross section 
modifications are referred to in terms of percentage, so that a 6\% 
increase is referred to as 106\%. Five packages are used in this 
analysis: {\sc fluka}{\footnotesize 105} (or {\sc fl}{\footnotesize 105}), 
{\sc fl}{\footnotesize 106}, {\sc fl}{\footnotesize 107}, 
{\sc gheisha}{\footnotesize 91} (or {\sc gh}{\footnotesize 91}) 
and {\sc gh}{\footnotesize 92}.

\subsubsection{Tuning the Hadron Cocktail Packages}\label{sssec:tuning}

To tune and validate a given hadron-cocktail package we can compare 
its output to three measured hadron distributions:

\begin{itemize}

\item The $\pt$ distribution of tracks that stop in 
MuID Gap2 (counting from 0), with $p_z$ larger than a given minimum 
value.

\item The $\pt$ distribution of tracks that stop in 
MuID Gap3 (counting from 0), with $p_z$ larger than a given minimum 
value.

\item The vertex $z$ distribution of reconstructed tracks, 
normalized to the collision-vertex distribution.

\end{itemize}

\begin{figure}
\includegraphics[width=1.0\linewidth]{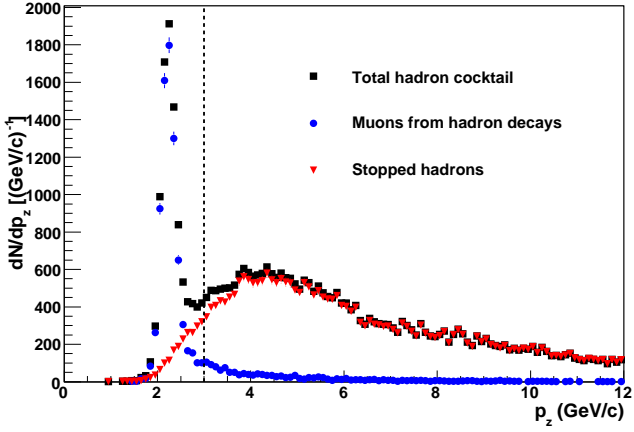}
\caption{(color online)
Simulated $p_{z}$ distributions for particles that stop in MuID 
Gap3: (Black squares) all particles; (red triangles) stopped 
hadrons; (blue circles) decay muons.}
\label{fig:pz-dist}
\end{figure}

Particles that stop in MuID Gap2 or Gap3 are those tracks for which 
no hit is found in the downstream gaps (Gap3 and/or Gap4). 
Figure~\ref{fig:pz-dist} shows the longitudinal-momentum ($p_{z}$) 
distribution of tracks stopping in MuID Gap3 obtained using a given 
hadronic cocktail. Decay muons are characterized by a sharp peak, 
corresponding to electromagnetic energy loss in the absorber 
material. Note that the same peak would be obtained for muons from 
heavy-flavor decay. In contrast, hadrons are characterized by a 
broad shoulder that extends to much larger values of $p_z$. For 
$p_z>p_z^{\rm min}$ (with $p_z^{\rm min}\approx 3$~GeV/$c$ in this 
example) one obtains a clean hadron sample. The hadron-input $\pt$ 
distributions can then be tuned so that a good match between the 
number of stopped hadrons in the simulation and in real data is 
achieved in each $\pt$ bin.

\begin{figure}
\includegraphics[width=1.0\linewidth]{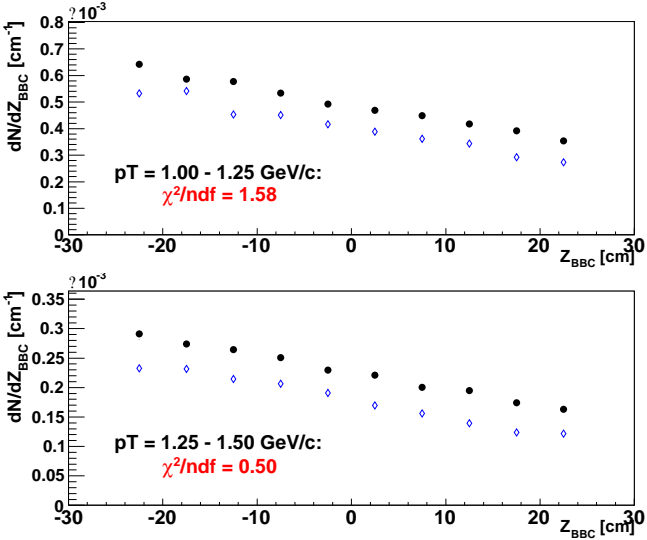}
\caption{(color online)
Vertex $z$ distribution of tracks reconstructed in North ($z>0$) 
MuID Gap4, for two transverse momentum bins. The real data (black 
closed circles) are compared to a given hadron-cocktail package 
(blue open diamonds). The offset between data and the hadron 
cocktail is the contribution from heavy-flavor decays.}
\label{fig:z-dist}
\end{figure}

Figure~\ref{fig:z-dist} shows, for two muon-$\pt$ ranges, 
comparisons for real data and hadron-cocktail simulations of the 
$z$-vertex distributions of $dN_\mu/dz_{\rm BBC}$ tracks, which (a) 
are reconstructed in the north muon arm (located at positive $z$), 
(b) reach the MuID Gap4, and (c) are normalized by the event vertex 
distribution $dN_{\rm evt}/dz_{\rm BBC}$.  The approximately linear 
dependence on $z_{\rm BBC}$ is entirely due to the contribution of 
muons from light hadrons decaying before the muon-tracker front 
absorber.  Muons from short-lived heavy-flavor hadrons have no 
measurable dependence on $z_{\rm BBC}$ and their contribution to 
the real-data sample is the source of the vertical offset between 
the hadron cocktail and the real-data distributions.  Therefore, 
the hadron-cocktail package can be tuned by matching the slopes of 
these two distributions in each $\pt$ bin.  The quality of this 
match is quantified by:

\begin{equation}
\chi_{\rm Gap4}^{2}(p_{T})=\sum_{i=1}^{N_{\rm bins}}\frac{(\Delta N_{i} - \overline{\Delta N})^{2}}{\sigma_{i}^{2}+\sigma_{mean}^{2}}
\end{equation}
where $N_{\rm bins}$ is the number of $z_{\rm BBC}$ bins; $\Delta N_{i}=dN_{I}/dz_{\rm BBC}-dN_{C}/dz_{\rm BBC}$ is the difference between the data and simulation for the $i^{th}$ $z_{\rm BBC}$ bin; $\overline{\Delta N}$ is the average difference over the entire $z_{\rm BBC}$ range; $\sigma_{i}$ and $\sigma_{mean}$ are the statistical uncertainties of $\Delta N_{i}$ and $\overline{\Delta N}$, respectively.  

\begin{figure}
\includegraphics[width=1.0\linewidth]{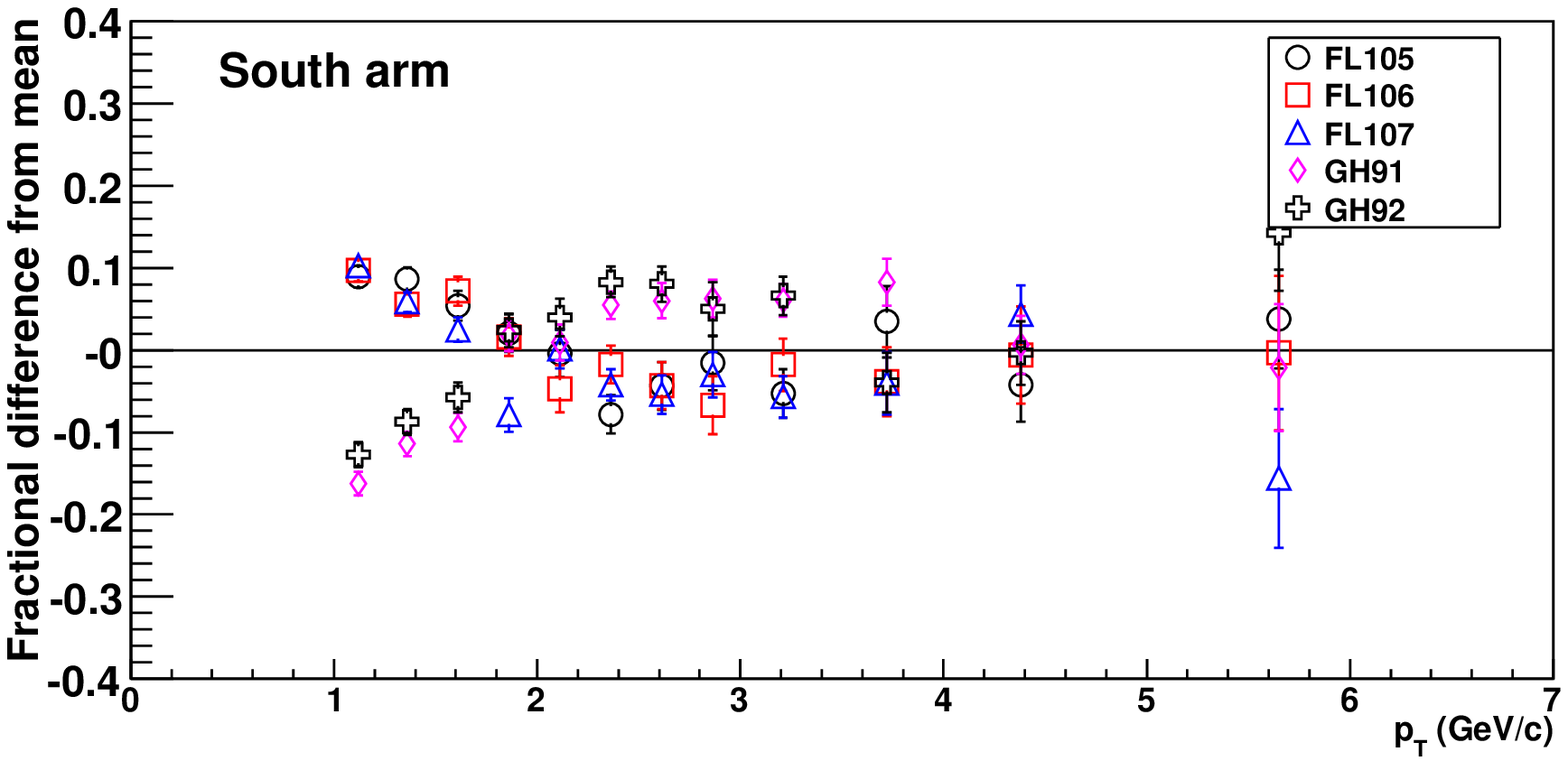}
\includegraphics*[width=1.0\linewidth]{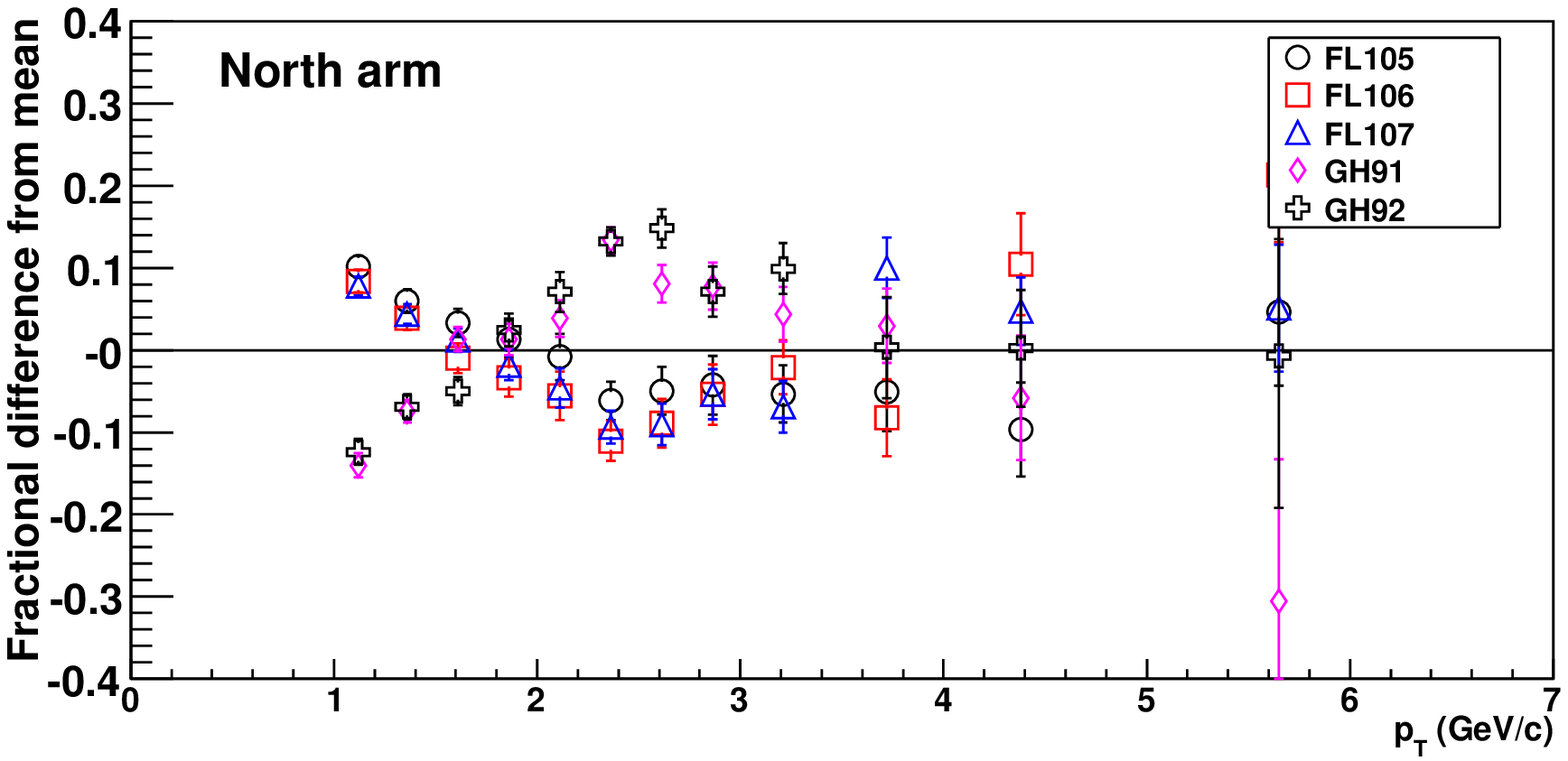}
\caption{(color online)
Relative dispersion between the $\nc$ yields obtained with 
the five hadron cocktails for the $\pp$ analysis. Each hadron 
cocktail package is compared to the mean of the five packages for 
the north (top panel) and south (bottom panel) muon arm.}
\label{fig:cocktailratios}
\end{figure}

Tuning of each hadron-cocktail package is achieved by iteratively 
selecting a set of $\pt$-dependent weights (applied to each track's 
thrown $\pt$) that simultaneously optimizes the agreement between 
data and simulation for the three distributions described above. 
Applying these weights to those simulated hadron tracks that reach 
MuID Gap4 determines the corresponding hadron contribution to the 
inclusive muon yield ($\nc$, Eq.~\ref{eq:smyield}).

Figure~\ref{fig:cocktailratios} shows the relative dispersion 
between $\nc$ values obtained for the five different hadron 
cocktail packages used for the $\pp$ analysis, as a function of 
$\pt$. For both muon arms, the largest differences exist between 
the {\sc gheisha} and {\sc fluka} cocktail packages for $\pt<2$\,GeV/$c$, with 
a spread of about 20\%. For $\pt>3$\,GeV/$c$, most of the 
dispersion between the packages is due to increased statistical 
uncertainty in the data yields used to tune the hadron cocktail.


\subsubsection{Systematic Uncertainties Associated with Individual 
Hadron Cocktail Packages}\label{ssec:cocktailSysErrors}

There are two systematic uncertainties associated with the 
implementation of a given hadron cocktail package:

\begin{description}

\item [$\mathbf\sigma_{\rm \bf SystPack}$] the uncertainty associated 
with the implementation of the hadron cocktail packages. It is 
comprised of two components: the uncertainty on the hadron cocktail 
input distributions and the so called MuID Gap3 to Gap4 matching 
uncertainty. The uncertainty on the hadron cocktail input 
distributions amounts up to 20\% and is correlated between the two 
arms. The uncertainty on the MuID Gap3 to Gap4 matching corresponds 
to tracks, in either real data or simulations, that get assigned an 
incorrect penetration depth, due to accidental addition of extra 
hits in the next MuID gap, or on the contrary, to detection 
inefficiencies. This uncertainty is evaluated using simulations. 
It is arm independent and amounts to 10\%. These two contributions 
are uncorrelated and added in quadrature.

\item [$\mathbf\sigma_{\rm \bf PackMismatch}$] the uncertainty 
that characterizes, as a function of $\pt$, the ability of a given 
hadron cocktail package to reproduce the measured distributions 
described in the previous section. To evaluate this uncertainty the 
cocktail is tuned three times, each time matching one of the three 
measured hadron distributions perfectly. The dispersion between the 
resulting background yields $\nc$ obtained with these three 
different tunings, along with the central value for $\nc$ obtained 
using the simultaneous tuning described above, is assigned to 
$\sigma_{\rm PackMismatch}$. A different value is calculated for 
each muon arm, each $\pt$ (and centrality) bin, and each of the 
five hadron cocktail packages. Mathematical details of the 
calculation are outlined in 
Section~\ref{ssec_central_point_determination}. Since the 
optimization is arm independent, this uncertainty is uncorrelated 
between the two muon arms. The magnitude of this uncertainty varies 
from 10 to 20\% depending on the muon arm and the $\pt$ bin.

\end{description}

\subsection{Other Background Sources}\label{ssec:otherbackground}

In addition to the hadronic background, other background sources 
include:

\begin{itemize}

\item muons from heavy-flavor-resonance leptonic decay (e.g. 
$\chic$, $\jpsi$, $\psiprime$ and the $\Upsilon$ family);

\item muons from Drell-Yan;

\item muons from light vector meson decay ($\rho$, $\phi$ and 
$\omega$).

\end{itemize}

These three sources contribute significantly less to the inclusive 
yields than the backgrounds from light hadrons. Monte Carlo 
simulations performed in the same manner as in~\cite{Adare:2011hf} 
show that their contribution to the final heavy-flavor muon $\pt$ 
spectrum is less than 5\% in the $\pt$ range used for this analysis 
and they have negligible impact with respect to the other sources 
of systematic uncertainties.

\subsection{\label{ssec_acc_eff}Acceptance and Efficiency Corrections}

Acceptance and efficiency corrections, $A\epsilon$, enter in the 
denominator of invariant yield measurements 
(Eq.~\ref{eq:smyield}). They are evaluated using simulated 
prompt single muons, propagated through the detector using the 
PHENIX {\sc geant}3 simulation and reconstructed with the same analysis 
code and the same track quality cuts as for the real data analysis. 
These corrections account for the detector's geometrical acceptance 
and inefficiencies (for example, due to tripped high voltage 
channels or dead front-end electronic channels). They also account 
for the muon triggers, reconstruction code and analysis cut 
inefficiencies.

A reference run, representative of a given data taking period, is 
chosen to define the detector's response to particles passing 
through it. This includes notably the list of inactive high-voltage 
and electronic channels. Remaining run-to-run variations with 
respect to this reference run are small due to the overall 
stability of the detector's performance, and are included in the 
systematic uncertainties ($\sigma_{\rm run to run} = 2\%$).

A comparison between the hit distributions in the MuTr and the MuID 
obtained for the reference run in real data and simulations is used 
to assign an additional systematic error on our ability to 
reproduce the real detector's response in the simulations. Areas 
with unacceptable discrepancies are removed from both the 
simulations and the real data using fiducial cuts. Remaining 
discrepancies are accounted for with a 8\% systematic uncertainty 
for the MuTr and 4.5\% for the MuID.

The hit multiplicity in the MuTr for $\cucu$ collisions is much 
higher than for $\pp$ collisions and for the single muon 
simulations. To account for deterioration of the reconstruction 
efficiency in presence of such high multiplicity events, simulated 
single muon events are embedded into real data $\cucu$ events 
before running the reconstruction and evaluating the $A\epsilon$ 
correction.

Another systematic uncertainty, $\sigma_{\rm p-scale}$, is assigned 
to a possible systematic bias between the particle's reconstructed 
momentum and its real momentum. This uncertainty is estimated by 
comparing the measured $\jpsi$ invariant mass (using the dimuon 
invariant mass distribution) and its Particles Data Group (PDG) 
value. This uncertainty amounts to $\sim 1.5$\%.

Table~\ref{tab:acc_eff_uncert} summarizes the acceptance and 
efficiency related uncertainties, which sum quadratically to 9.5\%.

\begin{table}[tbp]
\caption{Uncertainties in the acceptance and efficiency 
corrections. Individual components are added in quadrature to 
obtain the total value of $\sigma_{A\epsilon}$. }
\begin{ruledtabular}
\begin{tabular}{clc}
\multicolumn{2}{c}{Component} & Value \\
\hline
$\sigma_{\rm MuTr}$ & MuTr station data/MC  & 8\% \\
$\sigma_{\rm MuID}$ & MuID Gap4 efficiency uncertainty & 4.5\% \\
$\sigma_{\rm run\;to\;run}$ &Run to run variation & 2\% \\
$\sigma_{\rm p-scale}$ & momentum scale &  1.5\% \\
\\
$\sigma_{A\epsilon}$ & Total & 9.5\% \\
\end{tabular}
\end{ruledtabular}
\label{tab:acc_eff_uncert}
\end{table}

\subsection{\label{ssec_syst_uncert}Systematic Uncertainties}

This section summarizes systematic uncertainties associated with 
this analysis, most of which have been described in previous 
sections:

\begin{itemize}

\item Systematic uncertainties associated with individual hadron 
cocktail packages, $\sigma_{\rm SystPack}$ and $\sigma_{\rm 
PackMismatch}$ (Section~\ref{ssec:cocktailSysErrors});

\item Systematic uncertainty resulting from the dispersion of the 
results obtained with the different hadron cocktail packages 
(introduced in Section~\ref{sssec:packages}, mathematical details 
in Section~\ref{ssec_central_point_determination});

\item Systematic uncertainty on the acceptance and efficiency 
correction factors, $\sigma_{A\epsilon}$ (Section~\ref{ssec_acc_eff} and 
Table~\ref{tab:acc_eff_uncert});

\end{itemize} 

These systematic uncertainties are calculated independently for 
each arm, $\pt$ bin and centrality bin.

The first three uncertainties listed above (first two items) are 
related to the hadronic background estimate and are combined to 
form a $\sigma_{\rm model}$ systematic uncertainty, following a 
method described in Section~\ref{ssec_central_point_determination}.

For invariant cross section measurements (in $\pp$ collisions) and 
measurements of $\raa$ one must add to the 
uncertainties above the systematic uncertainty on the $\pp$ 
inelastic cross section seen by the minimum bias trigger, 
$\sigma^{pp}_{\rm BBC} = 9.6$\%. For $\raa$
measurements, one must also add the systematic uncertainty on the 
mean number of binary collisions ($\ncol$) in each centrality bin, 
as provided by the Glauber calculation used to determine this 
quantity.

Table~\ref{tab:sys_uncert} summarizes the systematic uncertainties 
in this analysis.

\begin{table}
\caption{Uncertainties in the single muon analysis. The individual 
components contribute to the final uncertainty as discussed in 
Section~\ref{ssec_central_point_determination}. }
\begin{ruledtabular}
\begin{tabular}{ccc}
 & Component & Value \\
\hline
 $\sigma_{\rm PackMismatch}$ & Package mismatch & varies, $\sim$10\%\\
 $\sigma_{\rm SystPack}$ & Single package uncertainty & varies, 10 - 20\%\\
 $\sigma_{A\epsilon}$ & Acceptance and efficiency & 9.5\% \\
$\sigma^{pp}_{\rm BBC}$ & $\epsilon_{\rm BBC}$ & 9.6\%\\
 $\sigma_{\ncol}$ & $\ncol$ & varies, 10 - 13\%\\
\end{tabular}
\end{ruledtabular}
\label{tab:sys_uncert}
\end{table}


\subsection{\label{ssec_central_point_determination}Determination 
of the Central Value for Heavy-Flavor-Muon Production Yields and 
$\raa$}

This section details the procedure used to combine the results from 
multiple hadron cocktail packages to obtain the central values for 
the $\pt$ spectra and $\raa$ and to propagate associated systematic 
uncertainties. This discussion includes the definition of 
$\sigma_{\rm PackMismatch}$ and $\sigma_{\rm Model}$. Throughout 
this section the variable $Q$ is used to represent either the 
invariant yield or $\raa$, for a given $\pt$ and centrality bin; 
the procedure is the same for both, except where noted explicitly.

\begin{enumerate}

\item For each $\pt$ bin $i$, hadron cocktail package $j$, 
and package tuning $k$, we calculate the value $Q_{i,j,k}$ where:

\begin{description}

\item [$k$=1] is the {\em optimal} tuning that best 
matches all three hadron distributions simultaneously 
(see Section~\ref{sssec:tuning});

\item [$k$=2] is the tuning that best reproduces the $\pt$ 
distribution of particles stopping in MuID Gap2;

\item [$k$=3] is the tuning that best reproduces the $\pt$ 
distribution of particles stopping in MuID Gap3; and

\item [$k$=4] represents the tuning that best reproduces the vertex 
$z$ distribution of particles reaching MuID Gap4.

\end{description}

The tuning $k=1$ is used for the central value whereas the other 
tunings are used to establish the systematic uncertainty for a 
single hadron cocktail package due to its inability to completely 
describe measured hadron distributions.

\item The package mismatch contribution to the uncertainty 
on the measurement $Q_{i,j,k}$ is estimated by the standard 
deviation between the four tunings, $k$:

\begin{equation}
\sigma^{2}_{{\rm PackMismatch},i,j}=\frac{1}{4} 
\sum^{4}_{k=1}(Q_{i,j,k} - \avrg{Q_{i,j,k}})^{2}
\end{equation}

\item For each $\pt$ bin $i$ and package $j$, the 
associated total uncertainty $\sigma_{i,j}$ is computed:

\begin{equation}
\begin{array}{rcl} 
\sigma^{2}_{i,j}&=&\sigma^{2}_{{\rm StatData},i} + \sigma^{2}_{{\rm StatPack},i,j} \\ 
&+&\sigma^{2}_{{\rm SystPack},i}+\sigma^{2}_{{\rm PackMismatch},i,j}\\
&+&\sigma^{2}_{A\epsilon,i},
\end{array}
\end{equation}
where the first two contributions, $\sigma^{2}_{{\rm StatData},i}$ 
and $\sigma^{2}_{{\rm StatPack},i}$ are the statistical 
uncertainties on the data and on the simulation and all other terms 
have already been introduced in previous sections.

\item Using $\sigma_{i,j}$ from step 3 we calculate the 
weighted mean of the $Q_{i,j}$ values obtained for the optimal 
tuning ($k=1$) of the different packages, $j$, in each $\pt$ bin, 
$i$:

\begin{equation} \nonumber
\avrg{Q_{i}}=\sum_{j=1}^{5}w_{i,j}Q_{i,j,k=1} 
\end{equation}
where
\begin{equation}\label{eq:weights}
w_{i,j} \equiv \frac{1/\sigma^{2}_{i,j}}{\displaystyle\sum_{j=1}^{5} 1/\sigma_{i,j}^{2}}.
\end{equation}

\item The total uncertainty on the final measurement is the 
variance of the weighted mean:
\begin{equation}
\begin{array}{rcl}
{\rm Var}(\avrg{Q_{i}})&=&\displaystyle\sum^{5}_{j=1} w^{2}_{i,j}\sigma_{i,j}^{2} \\
&+&2\displaystyle\sum^{5}_{j<m}w_{i,j}w_{i,m}\sigma^{2}_{{\rm common},i}
\label{eq:weightedError}
\end{array}
\end{equation}
where $\sigma_{{\rm common},i}$ is the part of the total 
uncertainty that is correlated between different packages:

\begin{equation}
\begin{array}{rcl}
\sigma^{2}_{{\rm common},i} &\equiv&\sigma^{2}_{{\rm StatData},i}  \\ 
&+&\sigma^{2}_{{\rm SystPack},i}+\sigma^{2}_{A\epsilon,i}
\end{array}
\end{equation}


\item For convenience, the total uncertainty ${\rm 
Var}(\avrg{Q_{i}})^{1/2}$ is split into a statistical uncertainty, 
a model-related systematic uncertainty and an acceptance and 
efficiency correction related systematic uncertainty:

\begin{equation}
{\rm Var}(\avrg{Q_{i}}) = \sigma_{{\rm StatCombined},i}^2 + \sigma_{{\rm model}, i}^2  + \sigma_{A\epsilon,i}^2
\label{eq:armError}
\end{equation}
with:
\begin{equation}
\sigma_{{\rm StatCombined},i}^2=\sigma^{2}_{{\rm StatData},i} + \frac{1}{5}\sum_{j=1}^5\sigma^{2}_{{\rm StatPack},i,j}
\end{equation}
and (by construction):
\begin{equation}
\sigma_{{\rm model},i}^2\equiv{\rm Var}(\avrg{Q_{i}}) - \sigma_{{\rm StatCombined},i}^2 - \sigma_{A\epsilon,i}^2
\end{equation}
so that the final measurement, in a given muon arm, is written:
\begin{equation}
\avrg{Q_{i}}\pm\sigma_{{\rm StatCombined},i}\pm\sigma_{{\rm model},i}\pm\sigma_{A\epsilon,i}
\end{equation}

\item The independent North and South measurements are combined using:
\begin{equation}
\avrg{Q_{i}}=\sum^{2}_{j=1} w_{i,j} Q_{i,j},
\end{equation}
where $i$ is the index of the $\pt$ bin, $j$ the arm index and 
$w_{i,j}$ a weight calculated in the same manner as in 
Eq.~\ref{eq:weights}, using the following total uncertainty on 
the measurement $Q_{i,j}$:
\begin{equation}
\sigma_{i,j}^2=\sigma_{{\rm StatCombined},i,j}^2 + \sigma_{{\rm 
model}, i,j}^2+\sigma_{A\epsilon,i,j}^2,
\end{equation}
which is identical to the expression of Eq.~\ref{eq:armError}, 
but explicitly includes the arm index, $j$.

The total uncertainty on the arm-averaged $Q_i$ value is calculated 
in a manner similar to Eq.~\ref{eq:weightedError}:
\begin{equation}
\begin{array}{rcl}
{\rm Var}(\avrg{Q_{i}})&=&\displaystyle\sum^{2}_{j=1}w^{2}_{i,j}\sigma_{i,j}^{2} \\
&+&2\displaystyle\sum^{2}_{j<m}w_{i,j}w_{i,m} \sigma^{2}_{{\rm arm\;common},i}
\end{array}
\end{equation}
where $\sigma^{2}_{{\rm arm\;common},i}$ is the systematic 
uncertainty common to both muon arms due to uncertainty on cocktail 
input.

For convenience, this uncertainty is again split into a statistical 
contribution $\sigma_{{\rm StatCombined},i}$ and a systematic 
contribution $\sigma_{{\rm SystCombined},i}$ defined by:
\begin{equation}
\sigma^{2}_{{\rm SystCombined},i}\equiv{\rm Var}( \avrg{Q_{i}})-\sigma^{2}_{{\rm StatCombined},i}
\end{equation}
so that the final, arm-averaged, measurement of $Q_i$ is written:
\begin{equation}
\avrg{Q_i}\pm\sigma_{{\rm SystCombined},i}\pm\sigma_{{\rm StatCombined},i}
\end{equation}

\end{enumerate}

As already noted in Section~\ref{ssec_syst_uncert}, for invariant 
cross section measurements (in $\pp$ collisions) and $\raa$
measurements one must add the systematic 
uncertainty on the $\pp$ inelastic cross section seen by the 
minimum bias trigger, $\sigma^{pp}_{\rm BBC}$ in quadrature to the 
uncertainties above. For $\raa$ measurements 
one must also add the systematic uncertainty on the mean number of 
binary collisions $\ncol$ in each centrality bin.

\section{Results}\label{sec:results}

\subsection{Heavy-Flavor Muon $\pt$ Distributions in $\pp$ and 
$\cucu$ Collisions}\label{ssec:ppspectrum}

\begin{figure}
\includegraphics[width=1.0\linewidth]{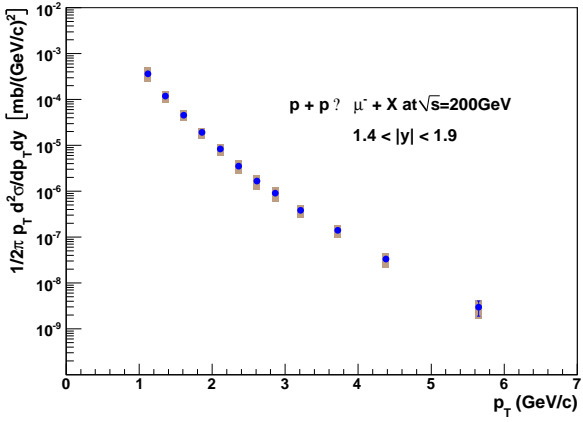}
\caption{(color online)
Production cross section of negative 
muons from heavy-flavor mesons decay as a function of $\pt$
in $p+p$ collisions at $\sqrt{s}=200$ GeV. }
\label{fig:ppsignal} 
\end{figure}

Figure \ref{fig:ppsignal} shows the production cross section of 
negatively charged muons from decays of open-heavy-flavor mesons as 
a function of $\pt$.  Vertical bars 
correspond to statistical uncertainties. Boxes correspond to the 
systematic uncertainties calculated following the steps described 
in Section~\ref{ssec_central_point_determination}. The measurements 
from both muon arms have been combined to reduce the overall 
uncertainty. Measured values for each $\pt$ bin are listed in the 
Appendix (Table~\ref{tab:inv_pp}).


\begin{figure}
\includegraphics[width=1.08\linewidth]{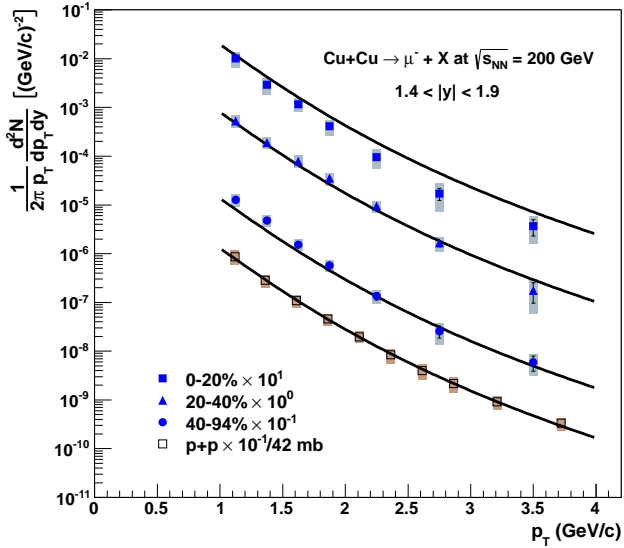}
\caption{(color online)
Invariant production yields of negative muons from 
heavy-flavor-mesons decay as a function $\pt$ in 
$p+p$ collisions at $\sqrt{s}=200$ GeV (open squares) and in 
$\cucu$ collisions for three different centrality intervals 
($40--94$\%, $20--40$\% and $0--20$\%), scaled by powers of ten for 
clarity (filled circles). The solid line associated to each set of 
points corresponds to a fit to the $\pp$ invariant yield 
distribution described in the text, scaled by the appropriate 
number of binary collisions $\ncol$ when comparing to the $\cucu$ 
measurements.
}
\label{fig:cucusignal}
\end{figure}

Figure~\ref{fig:cucusignal} shows the invariant yield of negative 
muons from heavy-flavor mesons decay for all analyzed $\cucu$ 
centrality classes, compared to the invariant yield measured in 
$\pp$ collisions. The solid lines correspond to a fit to the $\pp$ 
data using the function \mbox{$A(1+(\pt/B)^{2})^{-4.2}$}, similar 
to the one used in~\cite{Kaplan:1978}, scaled by the average number 
of binary collisions $\ncol$ for each $\cucu$ centrality bin. For 
peripheral ($40--94$\% centrality) and midcentral ($20--40$\% 
centrality) $\cucu$ collisions, a reasonable agreement is observed 
between the measurement and the scaled fit to the $\pp$ data, 
whereas for central collisions ($0--20$\%), a systematic difference 
is visible for high $\pt$ muons ($\pt\ge2$GeV/$c$), and the 
measurements are below the scaled $\pp$ fit, indicating a 
suppression of the heavy-flavor yields with respect to binary 
collision scaling, which is best quantified by measuring $\raa$ (see 
Section~\ref{ssec:RAA_result}).

\subsection{\label{ssec_charm_cross_results}Charm Cross Section, $\left.d\sigma_{c\bar{c}}/dy \right|_{\avrg{y}=1.65}$ in $\pp$ Collisions}

The $p+p$ heavy-flavor muon $\pt$ distribution is used to estimate 
the charm differential production cross section, 
$d\sigma_{c\bar{c}}/dy$ at forward rapidity ($\avrg{y}=1.65$), as 
described in detail in reference~\cite{Donnys_thesis}. The muon 
$\pt$ spectrum measured in $\pp$ collisions spans from $\pt=1$ to 
$7$~GeV/$c$. Estimation of the full charm charm cross section 
requires a theoretical calculation in order to extrapolate the 
measurement down to $\pt=0$ GeV/$c$. A set of 
fixed-order-plus-next-to-leading-log (FONLL) 
\cite{fonll:1998,fonll:2001} 
calculations have been used in this analysis.

The charm production cross section $d\sigma_{c\bar{c}}/dy$ is 
derived from the heavy-flavor muon cross section using:
\begin{equation}
d\sigma_{c\bar{c}}/dy = 
\frac{1}{BR(c \rightarrow \mu)} \cdot \frac{1}{C_{l/D}} 
\cdot \frac{d\sigma_{\mu^-}}{dy}
\end{equation}
where $BR(c \rightarrow D)$ is the total muon branching ratio of 
charm and is fixed to 0.103 in FONLL; $C_{l/D}$ is a kinematic 
correction factor, also provided by the FONLL calculation, which 
accounts for the difference in rapidity distributions between 
leptons and $D$ mesons; $d\sigma_{\mu^-}/dy$ is the total cross 
section for negative muons from heavy-flavor mesons decay, 
integrated over $\pt$ and estimated by extrapolating our 
measurement down to $\pt=0$~GeV/$c$ using the FONLL calculation.


\subsubsection{Extrapolation of the Data for $\pt<1.0$ GeV/$c$}

Low $\pt$ muons dominate the integrated heavy-flavor muon 
cross section due to the power-law like behavior of the $\pt$ 
distribution (Fig.~\ref{fig:ppsignal}): according to the central 
value of the FONLL calculation, the integrated charm cross section 
for $\pt^{\mu}>1$ GeV/$c$ represents about $6$\% of the total. 
Additionally, the contribution of bottom quark decay to the 
heavy-flavor muon $\pt$ distribution becomes increasingly relevant 
for $\pt>4$ GeV/$c$, but has a negligible contribution to the 
integral and is ignored hereafter.

The measured spectral shape matches the calculated shape. 
Therefore, extrapolation of the measured heavy-flavor muon $\pt$ 
spectra down to $\pt=0$~GeV/$c$ using FONLL is given by:
\begin{equation}\label{eq:scale_results}
\left.d\sigma_{c\bar{c}}/dy \right|^{\rm PHENIX}= \left.d\sigma_{c\bar{c}}/dy \right|^{\rm FONLL}\alpha^{\rm FONLL}
\end{equation}
where $\alpha^{\rm FONLL}$ is a constant determined by fitting the 
central values of the FONLL $\pt$ distribution to the data for 
$\pt>1$~GeV/$c$. It amounts to 3.75, and is used in determining the 
central point for PHENIX muons shown in Fig.~\ref{fig:ppcross}.

\subsubsection{Systematic Uncertainties on $\left.d\sigma_{c\bar{c}}/dy\right|_{\avrg{y}=1.65}$}

The total systematic uncertainty assigned to 
$\left.d\sigma_{c\bar{c}}/dy\right|_{\avrg{y}=1.65}$ is a 
combination of experimental and theoretical uncertainties, added in 
quadrature.  The experimental systematic uncertainty on the 
integral above $\pt > 1$,GeV/$c$ is determined by the appropriate 
quadrature sum of the systematic uncertainties on the individual 
$\pt$ points. This uncertainty is up/down symmetric and is equal to 
$32$~\%.

The theoretical uncertainty for 
$\left.d\sigma_{c\bar{c}}/dy\right|_{\avrg{y}=1.65}$ originates 
from the FONLL uncertainties. The variation in the FONLL 
calculation are determined by variation of the factorization scale, 
$\mu_{F}$, the renormalization scale, $\mu_{R}$, and the charm 
quark mass. Other contributions, such as fragmentation and parton 
distribution functions are smaller and neglected in this analysis.

The FONLL upper and lower bounds obtained by varying the scales and 
the charm quark mass are treated as approximations for a one 
standard deviation systematic uncertainty. The ratio of the 
measured $\pt$ distributions for $\pt>1$~GeV/$c$ to the upper and 
lower FONLL bounds are fit independently to determine the 
corresponding two normalization factors. The difference between 
these two normalization factors is then used as a theoretical 
uncertainty. This uncertainty is asymmetric and amounts to 
$^{+29}_{-37}$~\%. These FONLL systematic uncertainties are 
consistent with those of a previous study \cite{Donnys_thesis}, 
which examined the different $\pt$ distributions obtained by 
varying the FONLL parameters, $1.3< M_{c}$[GeV/$c$]$< 1.7$, $0.5 < 
\mu_R/m_T < 2$, $0.5 < \mu_F/m_T < 2$, with $m_{T}$ representing 
transverse mass. The different predicted theoretical $\pt$ ranged 
within an envelope of $\pm$35\% relative to the central spectrum.


\subsubsection{Integrated Charm Production Cross Section at $\avrg{y}=1.65$ in $\pp$ collisions}\label{ssec:ppintegrated}
 
\begin{figure}
\includegraphics[width=1.0\linewidth]{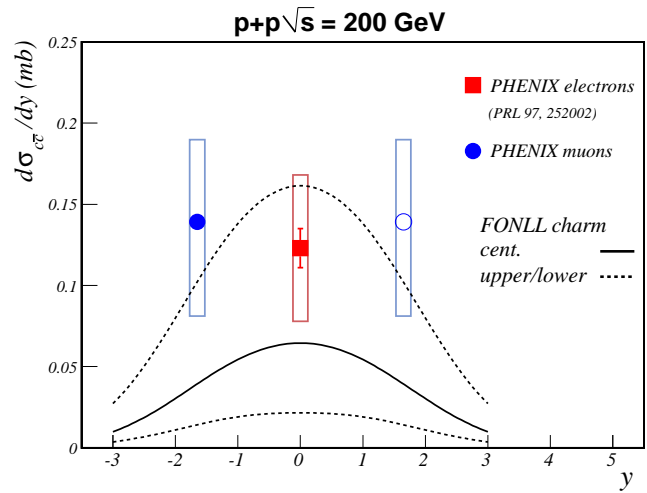}
\caption{(color online)
$c\overline{c}$ production cross section as a function of rapidity 
in $p+p$ collisions, measured using semileptonic decay to electrons 
(closed square) and to muons (closed circle).
}
\label{fig:ppcross}
\end{figure}

The integrated charm production cross section at forward rapidity 
($\avrg{y}=1.65$) obtained with this method is:

\begin{equation}
\left.d\sigma_{\ccbar}/dy\right|_{\avrg{y}=1.65} = 0.139\pm 0.029\ {\rm (stat)~}^{+0.051}_{-0.058}\ {\rm
(syst)}
\end{equation}
This measurement is shown in Fig.~\ref{fig:ppcross}, together 
with the measurement performed by PHENIX at 
midrapidity~\cite{Adare:2006hc}, as well as the FONLL calculation 
and its uncertainty band, calculated as discussed in the previous 
section. The full circle, located at $y=-1.65$, corresponds to the 
combined measurement performed in both muon arms. The open circle, 
located at $y=1.65$, corresponds to its mirror image.

\subsection{Heavy-Flavor-Muon $\raa$ in 
$\cucu$ Collisions as a Function of $\pt$}
\label{ssec:RAA_result}

\begin{figure}
\includegraphics[width=0.95\linewidth]{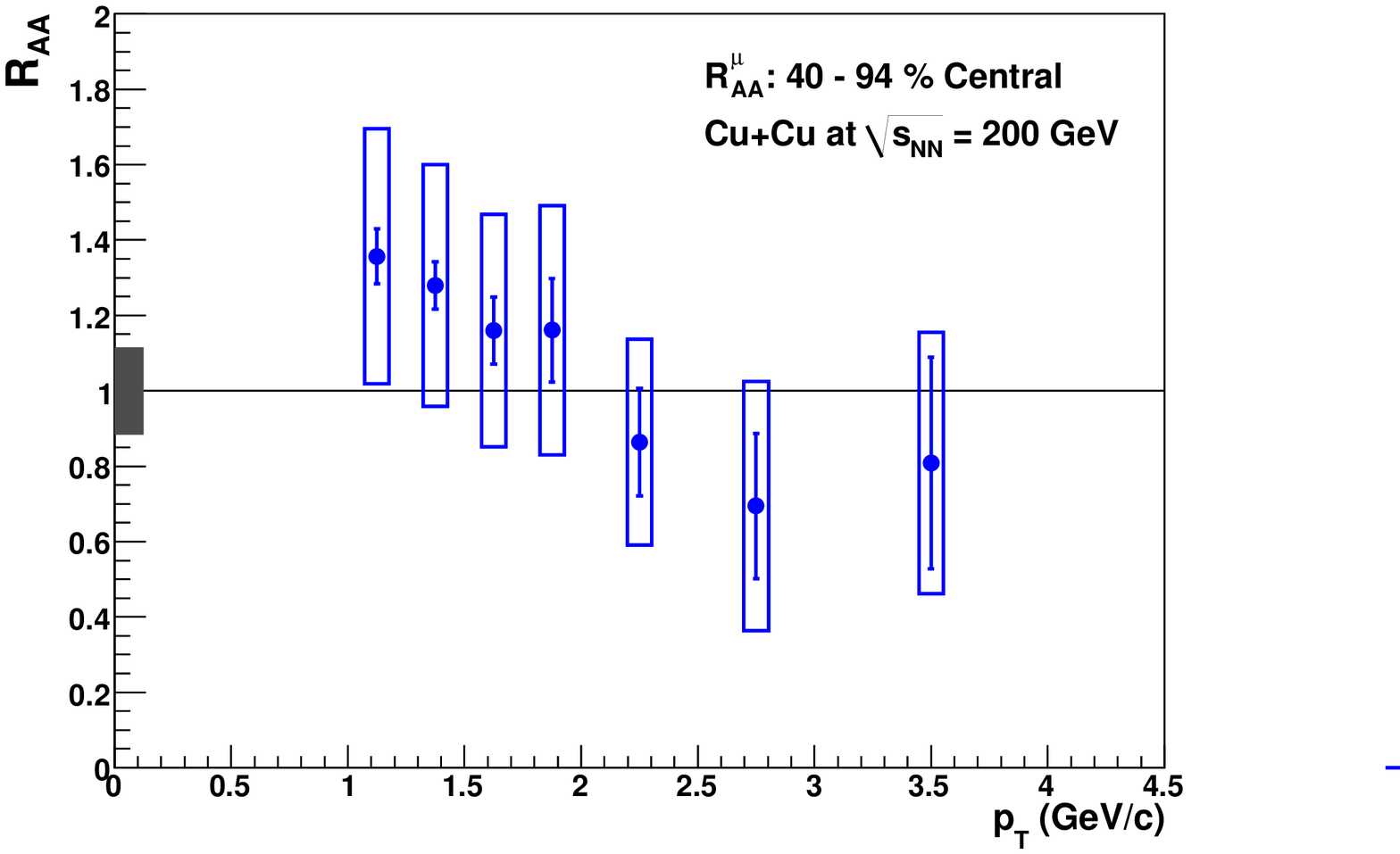} 
\includegraphics[width=0.95\linewidth]{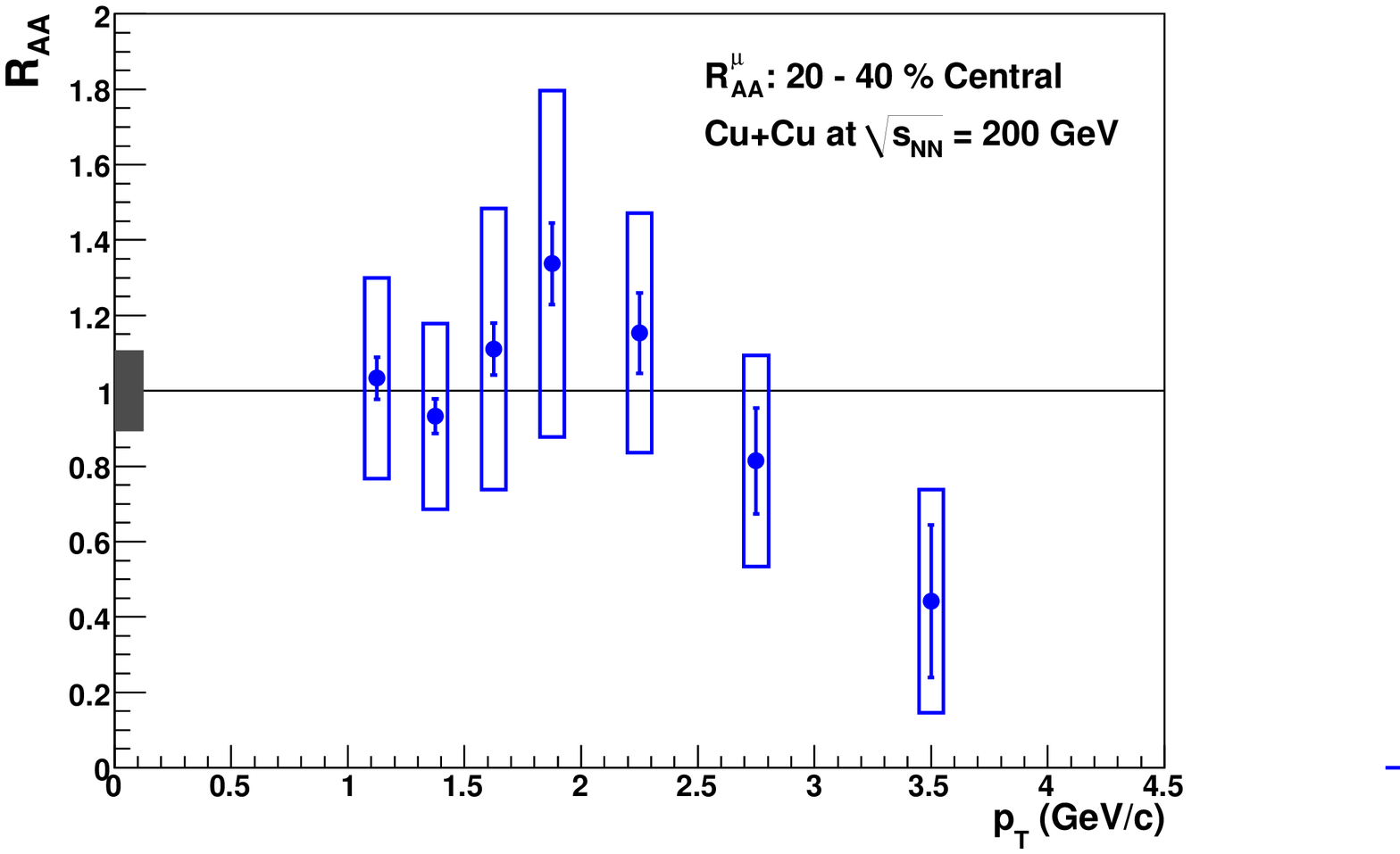} 
\includegraphics[width=0.95\linewidth]{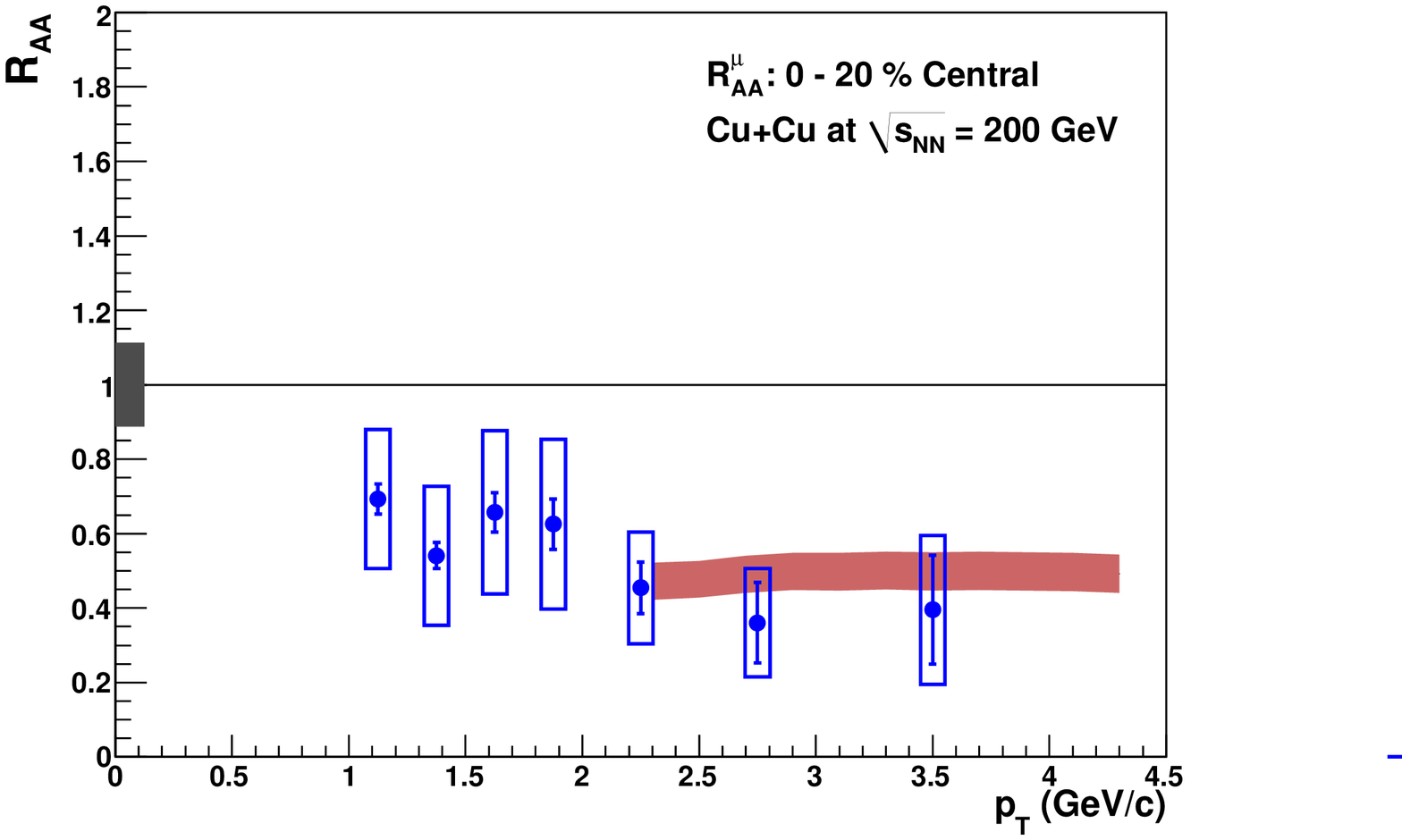}
\caption{(color online)
Transverse momentum distribution of $\raa$ 
for negative muons from heavy-flavor mesons decay in $\cucu$ 
collisions in the following centrality classes: 40--94\% (top 
panel), 20--40\% (middle panel) and 0--20\% (bottom panel). Also 
shown in the bottom panel is a theoretical calculation from 
\protect\cite{Vitev:private,Sharma:2009hn}, discussed in 
Section~\protect\ref{sec:discussion}}
\label{fig:raafinal}
\end{figure}

 \begin{figure}
  \includegraphics[width=1.0\linewidth]{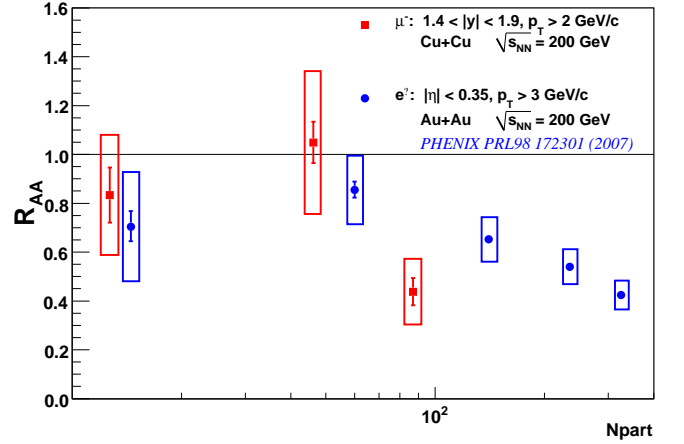}
\caption{(color online)
Comparison of $\raa$ as a function of 
$N_{\rm part}$ between heavy-flavor muons reconstructed at forward 
rapidity ($1.4<y<1.9$) and $\pt>2$GeV in $\cucu$ collisions (red 
squares) and nonphotonic electrons reconstructed at midrapidity 
and $\pt>3$GeV in $\auau$ collisions (blue circles).}
\label{fig:raa_muon_elec}
  \end{figure}

Figure~\ref{fig:raafinal} shows 
$R_{\rm AA}(\pt)$ for muons from heavy-flavor decay in $\cucu$ 
collisions as a function of muon $\pt$ for three 
centrality classes ($40--94$\%, $20--40$\% and $0--20$\%). As was 
the case for invariant yields and cross sections, the two independent 
measurements obtained with each muon arm are statistically 
combined, following the method discussed in 
Section~\ref{ssec_central_point_determination}. Vertical bars 
correspond to the statistical uncertainties; boxes centered on the 
data points correspond to point-to-point correlated uncertainties 
and the vertical gray band centered on unity corresponds to the 
uncertainty on $\ncol$, as listed in Table~\ref{table_centrality}. 
Also shown in the bottom panel of Fig.~\ref{fig:raafinal} is a 
theoretical calculation from \cite{Vitev:private,Sharma:2009hn}, 
discussed in Section~\ref{sec:discussion}. The measured values for 
each $\pt$ bin and each centrality class are listed in the Appendix 
(Table~\ref{tab:raa}).

\section{Discussion and Conclusions}
\label{sec:discussion}

The measurement of open-heavy-flavor muon production in $\pp$ 
collisions at $\sqrt{s}=200$~GeV reported in this paper is a 
significant improvement over the previous PHENIX published 
result~\cite{Adler:2006yu}. The transverse momentum range of the 
present measurement is extended to $\pt=7$ GeV/$c$ (compared to 
$\pt=3$ GeV/$c$ in the previous analysis). The differential 
production cross section is integrated over $\pt$ to calculate a 
production cross section at forward rapidity of 
$d\sigma_{c\bar{c}}/dy|_{\avrg{y} = 1.65} = 0.139\pm 0.029\ {\rm 
(stat)~}^{+0.051}_{-0.058}\ {\rm (syst)}$~mb. This cross section 
is compatible with a FONLL calculation within experimental and 
theoretical uncertainties. It is also compatible with expectations 
based on the corresponding midrapidity charm production cross 
section measured by PHENIX.

Muons from heavy-flavor decay have also been measured in $\cucu$ 
collisions at $\sqrt{s_{NN}} = 200$\,GeV/$c$, in the same rapidity 
and momentum range. This allows determination of the 
heavy-flavor-muon $\raa$ as a function of $\pt$
in three centrality classes, $40--94$\%, 
$20--40$\% and $0--20$\%. As shown in Fig.~\ref{fig:raafinal}, no 
suppression is observed across most of the transverse momentum 
range for muon yields measured in peripheral ($40--94$\%) and 
midcentral ($20--40$\%) $\cucu$ collisions compared to 
$\ncol$-scaled $\pp$ collisions. On the contrary, open heavy flavor 
production is significantly suppressed for central $\cucu$ 
collisions ($0--20$\%), with the largest effect observed for 
$\pt>2$~GeV/$c$. Interestingly, as demonstrated in 
Fig.~\ref{fig:raa_muon_elec}, the level of suppression for these 
higher $\pt$ heavy-flavor muons (the last red point on right) is 
comparable to the level of suppression observed for high $\pt$ 
nonphotonic electrons measured at midrapidity in the most central 
$\auau$ collisions (the last blue point on right). One expects the 
Bjorken energy density of the matter produced in the midrapidity 
region in the most central $\auau$ collisions to be at least twice 
as large as that of the matter produced in the forward rapidity 
region in most central $\cucu$ 
collisions~\cite{Cheuk-Yin:private,Iraklis_thesis}. Therefore the 
large suppression observed in $\cucu$ collisions suggests 
significant (cold) nuclear effects at forward rapidity in addition 
to effects due to strongly interacting partonic matter.

As shown in the bottom panel of Fig. \ref{fig:raafinal}, the 
suppression of open-heavy-flavor muon production for central 
$\cucu$ collisions is consistent with a recent theoretical 
calculation performed at the same rapidity ($y=1.65$) for 
$\pt>2.5$~GeV/$c$~\cite{Sharma:2009hn,Vitev:private}. This 
calculation includes effects of heavy-quark energy loss (both 
elastic and inelastic) and in-medium heavy meson dissociation. 
Additionally, the calculation accounts for cold nuclear matter 
effects relevant for open heavy flavor production 
\cite{Vitev_cold:2007}, namely shadowing (nuclear modification of 
the parton distribution functions of the nucleon) and initial state 
energy loss due to multiple scattering of incoming partons before 
they interact to form the $\ccbar$ pair.

New PHENIX inner silicon vertex detectors will greatly improve 
heavy flavor production measurements and allow separation of charm 
and bottom contributions~\cite{vtx,fvtx}.

\section*{ACKNOWLEDGMENTS}   

We thank the staff of the Collider-Accelerator and Physics
Departments at Brookhaven National Laboratory and the staff of
the other PHENIX participating institutions for their vital
contributions.  We acknowledge support from the 
Office of Nuclear Physics in the
Office of Science of the Department of Energy, the
National Science Foundation, Abilene Christian University
Research Council, Research Foundation of SUNY, and Dean of the
College of Arts and Sciences, Vanderbilt University (U.S.A),
Ministry of Education, Culture, Sports, Science, and Technology
and the Japan Society for the Promotion of Science (Japan),
Conselho Nacional de Desenvolvimento Cient\'{\i}fico e
Tecnol{\'o}gico and Funda\c c{\~a}o de Amparo {\`a} Pesquisa do
Estado de S{\~a}o Paulo (Brazil),
Natural Science Foundation of China (P.~R.~China),
Ministry of Education, Youth and Sports (Czech Republic),
Centre National de la Recherche Scientifique, Commissariat
{\`a} l'{\'E}nergie Atomique, and Institut National de Physique
Nucl{\'e}aire et de Physique des Particules (France),
Ministry of Industry, Science and Tekhnologies,
Bundesministerium f\"ur Bildung und Forschung, 
Deutscher Akademischer Austausch Dienst, 
and Alexander von Humboldt Stiftung (Germany),
Hungarian National Science Fund, OTKA (Hungary), 
Department of Atomic Energy (India), 
Israel Science Foundation (Israel), 
National Research Foundation and WCU program of the 
Ministry Education Science and Technology (Korea),
Ministry of Education and Science, Russian Academy of Sciences,
Federal Agency of Atomic Energy (Russia),
VR and the Wallenberg Foundation (Sweden), 
the U.S. Civilian Research and Development Foundation for the
Independent States of the Former Soviet Union, the US-Hungarian
NSF-OTKA-MTA, and the US-Israel Binational Science Foundation.

\section*{APPENDIX:~DATA TABLES}


Table~\ref{tab:inv_pp} gives the differential invariant cross 
section of muons from heavy-flavor decay in $\sqrt{s} = 200$~GeV 
$\pp$~collisions and corresponds to Fig.~\ref{fig:ppsignal}. 
Table~\ref{tab:raa} gives $\raa$ of muons 
from heavy-flavor mesons decay for the different centrality classes 
of $\snn=200$~GeV $\cucu$~collisions and corresponds 
to Fig.~\ref{fig:raafinal}.

\begingroup \squeezetable

\begin{table}[bh]
\caption{Differential-invariant cross section of negative muons 
from heavy-flavor mesons decay for 200 GeV $p+p$ collisions at 
midrapidity.\label{tab:inv_pp}}
\begin{ruledtabular} \begin{tabular}{ccccccccc}
$\pt$ (GeV/$c$)& $1/2\pi\pt\;d^2\sigma/d\pt d\eta$ (mb) & stat error &
syst error \\
\hline 
1.12 &3.64e-04 &1.55e-05 &1.23e-04 \\ 
1.36 &1.19e-04 &2.85e-06 &3.39e-05 \\ 
1.61 &4.57e-05 &1.15e-06 &1.25e-05 \\ 
1.86 &1.92e-05 &5.39e-07 &5.18e-06 \\ 
2.11 &8.31e-06 &3.02e-07 &2.27e-06 \\ 
2.36 &3.52e-06 &1.54e-07 &1.18e-06 \\ 
2.61 &1.67e-06 &9.21e-08 &5.74e-07 \\ 
2.86 &9.12e-07 &6.04e-08 &3.18e-07 \\ 
3.21 &3.83e-07 &2.28e-08 &1.22e-07 \\ 
3.72 &1.41e-07 &1.24e-08 &4.15e-08 \\ 
4.38 &3.34e-08 &3.49e-09 &1.12e-08 \\ 
5.65 &2.99e-09 &1.09e-09 &1.31e-09 \\ 
\end{tabular} \end{ruledtabular}

\caption{Nuclear-modification factor, $\raa$, of negative muons 
from heavy-flavor mesons decay as a function of $\pt$ for 
the specified centrality classes of $\cucu$~collisions at 
$\snn=200$~GeV.\label{tab:raa}}

\begin{ruledtabular} \begin{tabular}{ccccc}
Centrality & $\pt$ (GeV/$c$) & $\raa$ & stat error &
syst error \\
\hline 
&1.13 &6.93e-01 &3.98e-02 &1.87e-01 \\ 
&1.38 &5.41e-01 &3.49e-02 &1.87e-01 \\ 
&1.63 &6.57e-01 &5.32e-02 &2.20e-01 \\ 
0--20\%&1.875 &6.26e-01 &6.74e-02 &2.28e-01 \\ 
&2.25 &4.54e-01 &6.90e-02 &1.50e-01 \\ 
&2.75 &3.61e-01 &1.09e-01 &1.46e-01 \\ 
&3.5  &3.95e-01 &1.46e-01 &2.00e-01 \\
\\
&1.13 &1.03e+00 &5.59e-02 &2.66e-01 \\ 
&1.38 &9.32e-01 &4.63e-02 &2.46e-01 \\ 
&1.63 &1.11e+00 &6.95e-02 &3.72e-01 \\ 
20--40\%&1.875 &1.34e+00 &1.08e-01 &4.59e-01 \\ 
&2.25 &1.15e+00 &1.06e-01 &3.18e-01 \\ 
&2.75 &8.14e-01 &1.40e-01 &2.80e-01 \\ 
&3.5  &4.42e-01 &2.03e-01 &2.96e-01 \\
\\
&1.13 &1.36e+00 &7.27e-02 &3.38e-01 \\ 
&1.38 &1.28e+00 &6.26e-02 &3.21e-01 \\ 
&1.63 &1.16e+00 &8.87e-02 &3.08e-01 \\ 
40--94\%&1.875 &1.16e+00 &1.37e-01 &3.30e-01 \\ 
&2.25 &8.64e-01 &1.43e-01 &2.73e-01 \\ 
&2.75 &6.94e-01 &1.92e-01 &3.30e-01 \\ 
&3.5  &8.09e-01 &2.80e-01 &3.47e-01 \\
\end{tabular} \end{ruledtabular}
\end{table} 

\endgroup
   
\clearpage
   

\end{document}